\documentclass[nopreprintline,12pt]{elsarticle}

\usepackage{mathtools}
\usepackage[utf8]{inputenc}
\usepackage{hyperref}
\usepackage{graphicx}
\usepackage{wrapfig}
\usepackage[nodots]{numcompress}
\usepackage{multirow}  
\usepackage{pdfpages}
\usepackage{fancybox}
\usepackage{mdframed}

\begin{document}
\begin{frontmatter}

\title{FindZebra: A search engine for rare diseases\footnote{\textbf{NOTICE}: this is the author's version of a work that was accepted for publication in International Journal of Medical Informatics. Changes resulting from the publishing process, such as peer review, editing, corrections, structural formatting, and other quality control mechanisms may not be reflected in this document. Changes may have been made to this work since it was submitted for publication. A definitive version was subsequently published and is available online since 23 February 2013,  DOI:10.1016/j.bbr.2011.03.031.}}

\author[dtu,diku]{Radu Dragusin}
\author[dtu,findwise]{Paula Petcu}
\author[dtu,diku]{Christina Lioma}
\author[iva]{Birger Larsen}
\author[bisp]{Henrik L. Jørgensen}
\author[dtu,ucl]{Ingemar J. Cox}
\author[dtu]{Lars Kai Hansen}
\author[iva]{Peter Ingwersen}
\author[dtu]{Ole Winther\corref{cor}}
\ead{owi@imm.dtu.dk}

 \cortext[cor]{Corresponding author. Tel.: +45 4525 3895.}

\address[dtu]{DTU Compute, Technical University of Denmark, Denmark}
\address[diku]{Department of Computer Science, University of Copenhagen, Denmark}
\address[findwise]{Findwise, Copenhagen, Denmark}
\address[iva]{Information Systems and Interaction Design,\\ Royal School of Library and Information Science, Copenhagen, Denmark}
\address[bisp]{Department of Clinical Biochemistry, Bispebjerg Hospital, Copenhagen, Denmark}
\address[ucl]{Department of Computer Science, University College London, London, United Kingdom}

\begin{abstract}
\emph{Background:} The web has become a primary information resource about illnesses and treatments for both medical and non-medical users. Standard web search is by far the most common interface for such information. It is therefore of interest to find out how well web search engines work for diagnostic queries and what factors contribute to successes and failures. Among diseases, rare (or orphan) diseases represent an especially challenging and thus interesting class to diagnose as each is rare, diverse in symptoms and usually has scattered resources associated with it.\\
\emph{Methods:} We use an evaluation approach for web search engines for rare disease diagnosis which includes 56 real life diagnostic cases, state-of-the-art evaluation measures, and curated information resources. In addition, we introduce FindZebra, a specialized (vertical) rare disease search engine. FindZebra is powered by open source search technology and uses curated freely available online medical information.\\
\emph{Results:} FindZebra outperforms Google Search in both default setup and customised to the resources used by FindZebra. We extend FindZebra with specialized functionalities exploiting medical ontological information and UMLS medical concepts to demonstrate different ways of displaying the retrieved results to medical experts.\\
\emph{Conclusions:} Our results indicate that a specialized search engine can improve the diagnostic quality without compromising the ease of use of the currently widely popular web search engines. The proposed evaluation approach can be valuable for future development and benchmarking. The FindZebra search engine is available at \url{http://www.findzebra.com/}.
\end{abstract}


\end{frontmatter}

\section{Introduction}
\label{sec:introduction}
The web has become a primary source of information about illnesses or treatments \cite{cartright2011intentions}, with an exponential growth in both volume and amount of entries available \cite{hunter2006biomedical}. Important resources in locating medical information online are information retrieval systems, more commonly known as search engines. A December 2009 poll found that 66\% of web users have searched for medical information online \cite{purcell2010understanding}. This class of search activities, which goes beyond simple fact retrieval, is referred to as exploratory health search \cite{spink2004study,white2009cyberchondria}. It can be carried out by both expert and non-expert medical users.

A typical example of an expert medical user is a clinician. Diagnostic health search can also be seen as a coarse form of hypothetico-deductive reasoning \cite{cartright2011intentions}, where web search engines guide the iterative cycle of hypotheses about a disease being formulated from evidence, followed by the collection of additional discriminating evidence. According to recent studies, an increasing number of clinicians use web search engines to assist them in solving difficult medical cases, for instance when confronted with rare (or orphan) diseases \cite{bouwman2010doctor}. The exact definition of rare diseases in terms of prevalence threshold and requirement for severity varies across the globe, but a disease is, in general, said to be rare if it affects fewer than approximately one in two thousand individuals. A study\footnote{\url{http://archive.eurordis.org/article.php3?id article=454}.} conducted by the European Organisation for Rare Diseases (EURORDIS), showed that 40\% of rare disease patients were wrongly diagnosed before the correct diagnosis was given, and that 25\% of patients had diagnostic delays ranging between 5 and 30 years.

The current popularity of web search engines (primarily Google) and medical databases (primarily PubMed) for aiding in diagnosis may appear a bit surprising, as these tools are not optimised for this task. For example, a diagnostic query may be quite long, whereas web search engines are typically optimised for very short queries (2–3 terms long). Queries consist of lists of patient symptoms, often expressed as multi-word units. However, search engines often make term independence assumptions in order to increase efficiency. For instance, web search engines may not distinguish between “sleep deficiency, increased sexual appetite” and “sexual deficiency, increased sleep”, hence returning non-relevant results. Some symptoms listed in the clinician's query may not apply to the correct disease, and conversely, some pertinent symptoms for the correct disease may be missing from the query because they are masked under different conditions. However, search engines are designed to maximise the match between all the query terms and the returned documents.

In short, the clinicians' queries on rare diseases are likely to be more feature-rich but also harder for a search engine than ordinary web search queries, and should ideally be processed as such. Furthermore, the popularity-based metrics derived from hyperlinking (PageRank), user visit rates, or other forms of user recommendation that are commonly used by search engines are not likely to benefit the retrieval of rare diseases. These practices tend to favour webpages with many in-links (backlinks) or results often viewed by users (implicit feedback). Information on rare diseases, on the other hand, is generally likely to be very sparse and less hyperlinked than other medical content. Finally, efficiency concerns may lead to brute-force index pruning for web search, e.g. by removing from the index low frequency terms, or terms that are unusually long, such as “hydrochlorofluorocarbons” (\cite{manning2008introduction}, Chapter 5). Such practices may be particularly damaging for rare disease search, as the medical terminology involved may be exceptionally rare or formed by heavy term compounding. It is probably fair to conclude that familiarity, and the ease of use compared to traditional information search and diagnostic support systems (reviewed below) are the main factors contributing to the current popularity of general purpose web search engines in the clinical setting.

Motivated by these observations we asked to what degree can web search engines actually be used for rare disease diagnosis and what are the main contributing factors that determine success and failure. To try to answer these questions it is necessary to go through a number of steps. First of all an evaluation approach has to be set up. It should consist of cases of varying degrees of difficulty and retrieval performance measures to allow for quantitative comparisons between methods. Furthermore, the web search engine algorithms are not public so one can only to a limited degree change settings and thus interpret why a query returns a given set of results. Google offers a search engine customisation product called Google Custom Search Engine\footnote{\url{http://www.google.com/cse/}.} which has a few options for customisation that can be used to emphasise particular resources and thus determine how the choice of the information source (the index) influences the performance. If emphasising resources known to be authoritative in the rare disease domain improves the performance then one can conclude the huge index used by Google Web Search introduces noise. However, this will not give information along the “algorithm dimension”. We therefore made FindZebra, a search engine specifically designed to retrieve rare disease information for clinicians. It uses a specially curated dataset of rare disease information, which is crawled from freely available online authoritative resources. This means that FindZebra searches for rare disease information from a repository of “clean”, specialized resources, unlike web search engines that search the whole web and are hence likely to return spurious, commercial and less relevant results. The same index will be used for the customised versions of Google thus allowing us to gain an insight on the adequacy of the Google Search algorithm in rare disease diagnosis.

The rest of this article is organised as follows: \hyperref[sec:background]{Section \ref*{sec:background}} discusses background work on collecting and retrieving medical information automatically with a focus on rare disease data. \hyperref[sec:evaluation-approach]{Section \ref*{sec:evaluation-approach}} presents the evaluation approach. \hyperref[sec:findzebra]{Section \ref*{sec:findzebra}} presents our search engine, FindZebra and the information resources used for its index. \hyperref[sec:evaluation]{Section \ref*{sec:evaluation}} describes the evaluation, benchmarking FindZebra, different versions of Google Search, and PubMed against each other. \hyperref[sec:discussion]{Section \ref*{sec:discussion}} discusses the results and finally \hyperref[sec:conclusion]{Section \ref*{sec:conclusion}} summarises the findings of the paper.

\section{Background}
\label{sec:background}
Historically, the task of retrieving medical information has relied on authoritative resources, such as the 1879 Index Medicus (which ceased publication in 2004) \cite{hunter2006biomedical}. Since that time, the amount, availability and authority of medical resources have changed radically. More medical information is becoming freely available on the web; however, the authority of this information is not always easy to trace. A study by Eysenbach and Kohler \cite{eysenbach2002consumers} found that many users searching for medical information online largely ignored the credibility of web sites (in terms of source, design, scientific or official appearance, phraseology, and ease of use). Several researchers have also noted a need for improving the medical information on the web. Lewis’ \cite{lewis2006seeking} qualitative study into young peoples’ use of the web for medical information showed that users are often sceptical about the information they encounter. Nevertheless, there is evidence that young clinicians in particular are increasingly using the web to help them take medical decisions \cite{hughes2009junior}. Such findings demonstrate some of the conflicting opinions around the level and credibility of medical information on the web. In response to these issues, some organisations have initiated efforts to improve the general quality of medical information on the web, such as the Health on the Net Foundation,\footnote{\url{http://www.hon.ch}} or ways of augmenting web pages or search results with tools to support credibility assessment \cite{schwarz2011augmenting}.

Scepticism about the level and credibility of medical information on the web has motivated research in the direction of specialized medical web portals, repositories, database and search systems. The inadequacy of standard search technology for the task of medical retrieval has long been known. For instance, an earlier study \cite{hersh1998well} on how physicians use search systems to support clinical question answering and decision making revealed that search technology was inadequate for this purpose and generally retrieved less than half of the relevant articles on a given topic (a finding also supported by more recent studies, e.g. \cite{white2009cyberchondria}).

Furthermore, studies of expert users while they performed search tasks inside and outside their domains of expertise \cite{bhavnani2002domain} or using general purpose versus specialized medical search engines \cite{spink2004study}, identified domain-specific search strategies in each domain, and that such search knowledge is not automatically acquired from general-purpose search engines. Overall, the consensus seems to be that standard web search engines are not optimal for finding medical information online.

The retrieval of rare disease information is an even bigger problem, and efforts to address it date back to the 1990s. DXplain \cite{barnett1987evolving}, an early diagnostic support system that went online in 1996, was one of the first systems to display rare diseases separately from the rest. DXplain contained probabilities for over 4900 clinical manifestations associated with over 2200 unique diseases, yielding a total of over 230,000 unique disease interconnections. Another early effort was the London Dysmorphology Database, which contained information on rare dysmorphic syndromes and went online in 1999. This system provided the clinician with a manageable list of possible genetic syndromes (many of which are rare) for a particular case, with references, photographic information, and the possibility to register undiagnosed or unreported cases \cite{fryer1991london}. Further resources on rare diseases include the Online Mendelian Inheritance in Man (OMIM) database,\footnote{\url{http://www.ncbi.nlm.nih.gov/omim}} which specialises in human genes and genetic phenotypes and contains information for all Mendelian disorders and over 12,000 genes. Another major resource for rare diseases is the Orphanet database,\footnote{\url{http://www.orpha.net}} which contains information on more than 5900 rare diseases, and provides a service for retrieving data for rare diseases based on clinical signs or genes. Other databases on topics associated with rare diseases include POSSUMweb,\footnote{\url{http://www.possum.net.au}} which is a dysmorphology database that contains textual and photographic information on more than 3000 malformations, metabolic, teratogenic, chromosomal and skeletal syndromes. Phenomizer,\footnote{\url{http://compbio.charite.de/phenomizer}} a tool that uses the Human Phenotype Ontology (HPO), correlates phenotypic abnormalities with genetic disorders (OMIM entries) and contains around 9900 features and 5020 diseases. Furthermore, there are clinical decision support systems that aid in the diagnosis of one, or a few difficult to diagnose diseases \cite{tenorio2011artificial}, but their use is limited to verifying a diagnostic hypothesis, and the lack of standardisation hampers the integration of multiple such systems \cite{ahmadian2011role}.

The above are either diagnostic support or database systems, not search engines. The major difference between them is that database systems tend to process relational databases (well-structured data) whereas search engines tend to process unstructured data such as raw text or PDF files. The main data structure used in database systems is the relational table with well-defined values for each row and column. The main data structure used in search engines is the inverted index (index of {terms, document-IDs} entries) with a corresponding postings list (list of documents that contain a term) for each term. In most modern databases one can enable full-text search for text columns by building a type of inverted index and enabling Boolean or vector space search, effectively combining core database with search engine technology. See Chapter 10 in \cite{manning2008introduction} for more information on the fundamental differences between database systems and search engines in terms of retrieval model, data structures and query language.

The use of database systems for clinical diagnosis is not without problems. For instance, the search by clinical signs service provided by both Orphanet and Phenomizer is done using a controlled vocabulary (thesaurus). To search for a diagnosis in Orphanet, the user has to go through multiple steps. Going through a thesaurus and finding the right match can be a complex process that lengthens the diagnostic time, negatively impacts usability, and limits integration in the clinical environment. Similarly, in Phenomizer, the patient symptoms and signs must be selected from a predefined list compiled from the HPO ontology. Similarly, PubMed\footnote{\url{http://www.ncbi.nlm.nih.gov/pubmed}} is not a fully fledged IR system, but a medical database that comprises more than 21 million entries of biomedical literature from MEDLINE, life science journals, and online books. PubMed’s results are not ranked based on query relevance, but only on publication date, author name or other article meta-information that is not necessarily relevant in the search for a diagnosis. Moreover, when submitting a query without additional Boolean operators, only articles containing all query terms are retrieved, dramatically reducing the number of retrieved documents. A study of how medical experts use MEDLINE to gather evidence for clinical question-answering showed that users were only moderately successful \cite{hersh2002factors}.

Search engines with the exclusive purpose of retrieving information from specialized websites on rare diseases have been previously developed. For instance, the Rare Disease engine\footnote{\url{http://www.raredisease.org}} uses Google Custom Search Engine restricted to retrieve rare disease information. Another example is the Rare Disease Communities engine,\footnote{\url{http://www.rareconnect.org/en/search}} which aggregates search results from the eurordis.org, orpha.net, rarediseases.org and rarediseases.info.nih.gov websites. The attraction of medical search engines is that they are easy to use, fast, accessible, and their indices are continuously updated. While most of the medical database systems take as input complex structured queries requiring expert training, web search engines simply accept free-text queries. Moreover, medical database systems often return only results that exactly match the user query, whereas search engines may also use approximate matching algorithms. This is especially important for difficult cases where symptoms can be missing or misleading.
Despite the existence of specialized systems such as Orphanet, OMIM, Phenomizer or POSSUMweb, the general purpose Google Search is repeatedly mentioned in literature as a valuable tool for diagnosing difficult and rare disease cases \cite{bouwman2010doctor, lombardi2009search, tang2006googling, falagas2009pubmed}. Among the advantages of using Google in this setting are its comprehensive index, its ease of use, and the medical personnel's familiarity with it. Its main disadvantage in the scope of clinical diagnosis is that the results contain noise, with many being non-relevant (e.g. pages from non-authoritative sources such as forums and personal blogs, information on alternative medicine or sponsored content \cite{walji2005searching}).

The problem with general search engine algorithms in the context of clinical diagnosis is, as discussed above, that they are designed and optimised for web search. So even if popular, searching for diagnoses in Google or PubMed is still time-consuming; a specialized search engine could decrease search time and improve performance.

\section{Evaluation approach}
\label{sec:evaluation-approach}
The evaluation follows the standard paradigm of measuring functions of precision and recall at certain cut-off levels on a set of user queries \cite{croft2010search}. In the evaluation we want to address two properties of the different systems simultaneously, namely the quality of the dataset (the index) and quality of information retrieval algorithms for our particular task. We can to a large degree separate the two by using the Google Custom Search functionality. In this Section we describe the diagnostic queries, the curated rare disease index, the public web search engines (variants of Google Search) and database used (PubMed) together with the assessment and evaluation metrics. The description of our own search engine FindZebra is given in \hyperref[sec:findzebra]{Section \ref*{sec:findzebra}}.

\subsection{Diagnostic queries}
In total, 56 queries were used, which were created from difficult clinical cases, where the query text was extracted directly from the patient symptoms listed in the clinical cases. The composition of the 56 queries is as follows: 5 queries were created by a clinician (HLJ) on the basis of his expert knowledge; 25 queries were created from articles in the Orphanet Journal of Rare Diseases (OJRD) by RD and PP and curated by HLJ and 26 queries were taken from the British Medical Journal (BMJ) article of Tang and Ng \cite{tang2006googling}. All diagnoses except 4 from Tang and Ng are classified as rare. The full text and source of each query is included as an additional file to this article. The queries created from the Orphanet Journal of Rare Diseases are already indexed by Google so using this set poses a methodological problem as the source paper is likely to be highly ranked by Google Search. However, it turned out to be difficult to obtain “de novo” cases with a definite diagnosis so we opted for this approach to get a sufficiently large validation set and report “leave-source-out” result in these cases. The 26 queries of Tang and Ng are noticeably shorter, more vague and less realistic for medical professionals than the rest of our queries – however we include them in this experiment to facilitate comparison with the work of Tang and Ng.

\begin{table}
\begin{tabular}{|l|r|}
\hline 
\bf Resource & \bf Entries\\ 
\hline
\shortstack[l]{Online Mendelian Inheritance in Man (OMIM) \\ \url{http://www.ncbi.nlm.nih.gov/omim}} & 20,369\\
\hline
\shortstack[l]{Genetic and Rare Diseases Information Center (GARD) \\ \url{http://rarediseases.info.nih.gov/GARD}} & 4578\\
\hline
\shortstack[l]{Orphanet \\ \url{http://www.orpha.net}} & 2967\\ 
\hline
\shortstack[l]{Wikipedia \\ \url{http://www.wikipedia.org/}} & 2239\\
\hline
\shortstack[l]{National Organization for Rare Disorders (NORD) \\ \url{http://rarediseases.org}}  & 1230\\ 
\hline
\shortstack[l]{Genetics Home Reference \\ \url{http://ghr.nlm.nih.gov}} & 626\\
\hline
\shortstack[l]{Madisons Foundation Rare Paediatric Disease Database \\ \url{http://www.madisonsfoundation.org}} & 522\\
\hline
\shortstack[l]{About.com Rare Disease Database \\ \url{http://rarediseases.about.com}} & 316\\
\hline
\shortstack[l]{Health on the Net Foundation Rare Disease Database \\ \url{http://www.hon.ch}} & 183 \\
\hline
\shortstack[l]{Swedish National Board of Health and Welfare\\ \url{www.socialstyrelsen.se/rarediseases}} & 114\\
\hline
\end{tabular}
\caption{This table displays the resources used to compile the dataset of rare disease information used in this work.}
\label{table:dataset}
\end{table}

\subsection{The curated rare disease index}
\label{sec:index}
In order to create a high quality dataset of rare disease information, a number of authoritative, carefully curated medical resources were selected. Specifically, 33,144 documents were crawled from the resources shown in \hyperref[table:dataset]{Table \ref*{table:dataset}}. We estimate that this dataset covers well over 90\% of Orphanet’s list of rare diseases and more than 50\% when restricting to exact name matches. Resources maintained or curated by non-medical experts, such as blogs or support groups, were not included in the dataset. However, medically curated patient organisation resources, such as Madisons,\footnote{\url{http://www.madisonsfoundation.org}} were included. We chose not to include PubMed because it risked introducing too much irrelevant material, as we found no scalable and accurate way of selecting only those PubMed articles covering rare disease topics.

\subsection{Web search engines and database}
\label{sec:systems}
The publicly available search engines and database we tested our queries on are:
\begin{enumerate}[1.]
	\item Google Search\footnote{\url{http://www.google.com}. The details of it's ranking algorithm are not publicly known, however PageRank plays an important role, see Udi Manber, "\href{http://googleblog.blogspot.com/2008/05/introduction-to-google-search-quality.html}{Introduction to Google Search Quality}"}, which retrieves information from the web as indexed by Google.
	\item Google Custom Search set up to retrieve information from the web but emphasizes the sources of the curated rare disease index. We call this Google Custom in the following.
	\item Google Custom Search set up to retrieve information only from the sources of the curated rare disease index. We call this Google Restricted in the following.
	\item PubMed.
\end{enumerate}
The two Google Search Custom Search variants use the freely available Google Custom Search functionality. Note that Google imposes a limit on query length: any query longer than 32 terms is automatically truncated to the first 32 terms. Only one query in our collection is longer than 32 terms.

\subsection{Assessment and evaluation metrics}
The top 20 retrieved documents for each query were assessed by the authors as either relevant or non-relevant as follows:
\begin{itemize}
	\item relevant documents should address predominantly the correct disease in the title or within the first 400 words, and name it using any of its synonyms listed in Orphanet;
	\item in cases of inherited diseases, e.g. autosomal neonatal form of Adrenoleukodystrophy, documents treating the main disease, e.g. X-linked Adrenoleukodystrophy, are relevant;
	\item documents treating different types of the correct disease, e.g. Loeys–Dietz syndrome type 1A instead of Loeys–Dietz syndrome type II, are relevant;
	\item documents treating predominantly other diseases and mentioning the correct disease as an alternative diagnostic or pointing to it are not relevant;
	\item documents listing many diseases are not relevant if the correct disease is listed after the first 10.
\end{itemize}

Based on the above assessments, we computed the following evaluation metrics, all of which are standard in search engine evaluation: the average retrieval precision at rank $k$ ($P@k$) and the mean reciprocal rank (MRR). $P@k$ is the percentage of retrieved documents that are relevant after $k$ documents (whether relevant or non-relevant) have been retrieved, averaged over all queries:
\begin{equation}
P@k=\frac{1}{N}\sum_{i=1}^{N}\frac{\left | \left \{ Rel \right \} \right | \cap \left | \left \{ k \right \} \right |}{\left | \left \{ k \right \} \right |}
\end{equation}
where $N$ is the number of queries and $Rel$ the relevant documents for query $i$. This measure closely correlates with user satisfaction. However, it has been criticised because the constant cut-off of $k$ represents very different recall levels for different queries \cite{buckley2004retrieval}. For this reason, we also use MRR \cite{voorhees1999trec}, which corresponds to the multiplicative inverse of the rank of the first relevant document retrieved. Specifically, the reciprocal rank for a given query is the reciprocal of the rank position of the highest ranking relevant document for the query. MRR is the average of the reciprocal rank over queries:
\begin{equation}
\text{MRR}=\frac{1}{N}\sum_{i=1}^{N}\frac{1}{r_{i}}
\end{equation}
where $N$ is the number of queries and $r_i$ is the highest position of a relevant document for query $i$. This measure focuses on the retrieval quality of the very top of the ranked list.
In addition to the above measures, we also report the number of queries for which at least one relevant document is retrieved for ranks 1–10 and 1–20.

\section{FindZebra: a search engine for rare diseases}
\label{sec:findzebra}
Our search engine is called FindZebra, as zebra is a name often given to rare diseases by medical professionals \cite{haran2011clinical}. The interface of the search engine located at \url{findzebra.com} is very similar to that of standard web search engines so it should be straightforward to use by anyone familiar with web search. FindZebra is based on Indri \cite{strohman2005indri}, a state-of-the-art open source experimental information retrieval system. Specifically, we use Indri’s indexing and retrieval functions, on top of which we build an interface and several functionalities specifically tailored to rare disease diagnosis by clinicians. As corpus we use the curated rare disease resources described in Section 3.2. The retrieval time of our system was less than 0.5s per query on a virtual machine allocated with 1 GB RAM, on an Intel Xeon E5530 clocked at 2.40 GHz. Since the available information will change continuously, the index of the search engine will be updated every 3 months. The technical details of FindZebra's standard search functionality are described in \hyperref[sec:standardsearch]{Section \ref*{sec:standardsearch}} and the added functionality based upon UMLS medical concepts is described in \hyperref[sec:UMLSsearch]{Section \ref*{sec:UMLSsearch}}.

\subsection{Standard search}
\label{sec:standardsearch}
The system ranks documents decreasingly by their estimated relevance to the user query using the state-of-the-art query likelihood ranking model of the Language Model framework \cite{croft2003language} with Jelinek-Mercer and Dirichlet smoothing \cite{zhai2004study}. The respective equations for Jelinek-Mercer and Dirichlet smoothing are:
\begin{equation}
p\left ( q_{i} | D\right )=\prod_{i=1}^{N}\left ( 1-\lambda  \right )\frac{fq_{i},D}{\left | D \right |}+\lambda \frac{cq_{i}}{\left | C \right |}
\end{equation}
\begin{equation}
p\left ( q_{i} | D\right )=\prod_{i=1}^{N}\frac{fq_{i},D + \mu \tfrac{cq_{i}}{\left |C  \right |}}{\left | D \right |+\mu}
\end{equation}
where $p(q_i|D)$ is the probability of query term $q_i$ given document $D$, $fq_i,D$ is the frequency of $q_i$ in $D$, $cq_i$ is the frequency of $q_i$ in the collection of all documents, $|D|$ is the number of terms in document $D$, $|C|$ is the number of documents in the collection of all documents, $\lambda$ is the Jelinek-Mercer smoothing parameter ($0 \leq \lambda \leq 1$), and $\mu$ is the Dirichlet smoothing parameter. Parameters were set to default settings ($\mu = 2500$, $\mu = 0.9$, as described in \cite{zhai2004study}). These settings could be tuned in order to optimise the system's performance, for instance by ranging the parameter values across their respective ranges. We did not tune these parameters at this stage in order to avoid over-fitting our system's performance to our data.

This retrieval model performs basic text search without addressing term dependence and we chose it because it is the best-performing model in this category according to recent Text Retrieval Evaluation Conference (TREC) findings \cite{voorhees2011trec}. Documents are retrieved from the curated rare disease dataset described in \hyperref[sec:index]{Section \ref*{sec:index}} and displayed to the user in a simple interface. It is also possible to specify whether the documents should be retrieved from the whole rare and genetic disease dataset or only the rare disease resources of the dataset.

\subsection{Using the UMLS medical concepts in search}
\label{sec:UMLSsearch}
Motivated by the goal of facilitating medical diagnostic search, FindZebra also offers the options of (a) clustering the retrieved documents by UMLS medical concepts (diseases) derived from the document title, and (b) ranking UMLS concepts as opposed to documents. Both options aim to facilitate cases where the search engine retrieves several documents covering the same disease. The aim is to select and group these documents in flexible ways that, on the one hand, can facilitate a user's navigation through the retrieved results, and on the other hand, allow the display of a potentially more diverse set of results (which considers the top $j$ retrieved documents) than standard search (which is limited to the top $n$ retrieved results). We used $j=50$ and $n=20$ in FindZebra.

The mapping of documents to UMLS medical concepts was performed with MetaMap,\footnote{\url{http://metamap.nlm.nih.gov}. Subset of UMLS Metathesaurus 2011AA including 6 disease-related sources: ICD10CM, OMIM, Disease Database, DXP, QMR and RAM. The subset was extracted using \href{http://www.nlm.nih.gov/research/umls/implementation_resources/metamorphosys/help.html}{MetamorphoSys}, the UMLS customisation tool and uses \href{http://www.nlm.nih.gov/research/umls/knowledge_sources/metathesaurus/index.html}{UMLS Metathesaurus}} a standard tool of the US National Library of Medicine, which uses freely accessible medical ontologies and classifications recognised by the US National Institute of Health. We only select the maximum scoring mappings for each document. In order to achieve a near complete mapping of the articles we used the following three-step procedure. Allowing overmatches and truncate to mappings with matching scores above a certain threshold (600), 90\% of titles were matched to concepts. For the remaining unmapped titles we reduced the length of the OMIM titles to the first disease name variant and ran MetaMap with the same parameters. In the final step for the remaining unmapped articles, we ran MetaMap using a larger subset of the Metathesaurus, the datasets in Category 0. In the end 99.75\% of the titles were mapped to medical concepts.

FindZebra accepts UMLS concept identifiers as queries. In that case, documents are retrieved by matching the UMLS concept identifiers in the query to the UMLS concept identifiers that correspond to the document titles. This correspondence is already indexed in the system, and hence there is no extra delay at retrieval time.

FindZebra can also cluster the top $j=50$ retrieved documents according to their UMLS concepts. Clustering is performed by simply grouping together the retrieved documents associated with the same medical concept (i.e. disease) and using the highest ranking document to represent the cluster. If clicked, each cluster expands to reveal information on the documents it contains sorted by rank, thus allowing to zoom in on documents of interest. This option offers a quick summary of the main retrieved medical conditions.

The system can furthermore rank UMLS medical concepts directly. In that case, the search results consist of a list of UMLS medical concepts which, when clicked, point to their corresponding documents. We calculate the ranking score for UMLS concept $i$, $Ci$, by the following formula:
\begin{equation}
Score\left ( C_{i} \right )=\left | C_{i} \right |+\sum_{d \in C_{i}}^{\ }\frac{1}{R_{d}}
\end{equation}
where $|C_i|$ is the number of documents containing concept $C_i$, the sum goes over all documents containing the concept, and $R_d$ is the rank of the document according to the query. This alternative display is another succinct way of visualising the main medical concepts related to the user query. An example of the use of UMLS concepts in FindZebra is given in \hyperref[sec:UMLSexample]{Section \ref*{sec:UMLSexample}}.

\begin{table}
\begin{tabular}{|l|r|r|r|r|r|}
\hline 
\multirow{2}{*}{\bf Method} & \multirow{2}{*}{\bf MRR} & \multirow{2}{*}{\bf P@10} & \multirow{2}{*}{ \bf P@20} &\multicolumn{2}{|c|}{ \bf Unanswered queries}\\ 
& \bf & \bf & \bf  & \bf in top 10 & \bf in top 20 \\ 
\hline
FindZebra&0.385&0.125&0.089&35 (62.5\%)&38 (67.9\%)\\
Google Search&0.206&0.07&0.056&16 (28.6\%)&18 (32.1\%)\\
Google Custom &0.206&0.088&0.071&17 (30.4\%)&21 (37.5\%)\\
Google Restricted&0.098&0.013&0.006&6 (10.7\%)&6 (10.7\%)\\
PubMed&0.128&0.021&0.016&7 (12.5\%)&9 (16.07\%)\\
\hline
\end{tabular}
\caption{This table shows the performance of our system against three different versions of the Google Search engine in terms of retrieval precision at rank $k$ ($P@k$), the mean reciprocal rank (MRR) and the fraction of queries with the correct result in top $k$.}
\label{table:resultsoverview}
\end{table}

\section{Evaluation}
\label{sec:evaluation}
We evaluate and compare FindZebra and the four other systems presented in \hyperref[sec:systems]{Section \ref*{sec:systems}} from two perspectives. On the one hand following the standard paradigm of computing statistical measures of precision and recall, 
\begin{wrapfigure}{l}{0.30\textwidth}
\includegraphics[scale=0.54]{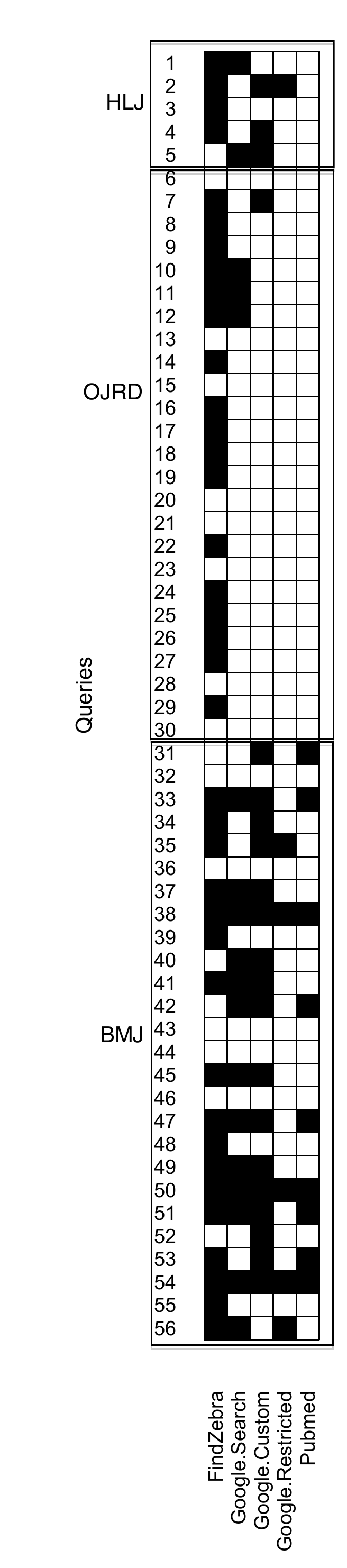}
\caption{An overview of the queries for each of the five systems, represented as a binary matrix.}
\label{fig:overview}
\end{wrapfigure}
a commonly used approach in evaluating information retrieval systems. \hyperref[table:resultsoverview]{Table \ref*{table:resultsoverview}} shows the retrieval precision at rank $k$ ($P@k$) and the mean reciprocal rank (MRR) of our experiments averaged for all 56 queries. The result for each query is given in Supplementary Table 2
\cite{de2010granulomatous,
bohm2010case,
schneider2010novel,
orsoni2010case,
al2010case,
acien2010hereditary,
poala2010case,
mejia2010genetic,
cassiman2009novel,
drera2009loeys,
ritelli2009arterial,
spoerl2009multisegmental,
marchiori2009pulmonary,
zamel2008abetalipoproteinemia,
virginie3carglumic,
joy2007alstrom,
ceruti2007successful,
gruhn2007successful,
al2007craniocervical,
sills2007further}. 
Secondly, we analyse the results returned for the queries by each of the systems in order to get a deeper understanding of the strength and weaknesses of each approach. This should also have particular value to clinicians.

As can be seen in \hyperref[fig:overview]{Figure \ref*{fig:overview}}, there is a clear distinction on how the systems perform overall, depending on the origin of the queries. Supplementary Table 3 summarizes the performance of all approaches in terms of queries for which the correct result appears in the top 20 ranks. The results show that all other systems except FindZebra have difficulty in addressing queries from OJRD. For these queries, FindZebra returns correct results for 17 of the 25 queries (68\%), whereas Google Search and Google Custom return correct results for 3 (12\%) and one (4\%), respectively. For the OJRD queries, neither Google Restricted, nor PubMed manage to return correct results for any of the queries. For the HLJ queries, FindZebra still has a lead, managing to return correct results for 4 of the 5 queries (80\%), with Google Search returning correct results for 2 (40\%), Google Custom returning correct results for 3 (60\%), Google Restricted returning correct results for one (20\%) and PubMed for none. For the BMJ queries, which were specifically devised for Google Search, the differences between systems are less pronounced. FindZebra and Google Custom are leading, both returning correct results for 17 of the 26 BMJ queries (65\%), with Google Search returning correct results for 13 (50\%), Google Restricted 5 (19\%) and PubMed 9 (35\%).

In 13 of the 56 queries, none of the systems was able to return the correct result. Of these, 8 were from OJRD and the remaining 5 were from BMJ. It is interesting to note that for both OJRD and BMJ queries, there have been comments from clinicians that some queries are particularly difficult to address. Specifically, HLJ identified 3 queries from OJRD (13, 20, and 21) where the symptoms are probably not specific enough to identify the correct cause. Additionally, Tang and Ng commented on the difficulty of 3 other queries (36, 39, and 43), noting that the first two are less likely to be successful because they cover a complex disease with non-specific symptoms, while the last one is less likely to be correctly addressed, because it covers a rare presentation of a common disease. Of these 6 queries labelled as particularly difficult, only FindZebra is able to find the correct result to one (query 39) at rank 8.

In 17 of the 56 queries, FindZebra is the only system with a correct result, with 13 of these queries from OJRD. Google Custom is the only system to return correct results for query 52. None of the other systems manages to be alone in returning correct results for any of the queries. There are 5 queries that are not addressed correctly by FindZebra, but for which one or more of the other systems return correct results. In fact, Google Custom returns correct results for all the 5 queries, with Google Search returning correct results for 3, and PubMed managing to return correct results for 2 of them.

\subsection{Use of UMLS medical concepts example}
\label{sec:UMLSexample}
The option of clustering documents by UMLS concepts is better suited for medical professionals who wish to quickly read about the correct diagnosis; instead of having to browse all the retrieved documents, they can focus on the cluster of documents that contains the correct diagnosis. Consider, for example, query 25 (see Supplementary Table 1), for which none of the Google Search options return a relevant result, and for which FindZebra standard search returns two relevant documents with the correct diagnosis at ranks 4 and 10. By selecting the option of clustering documents by UMLS medical concepts for this query, the correct diagnosis shows up as the main title of the third cluster. This cluster contains three documents on that disease that were originally at ranks 4, 10 and 27. Hence, whereas standard search retrieves two relevant documents at rank 4 at best, this clustering option retrieves three relevant documents at rank 3.

The option of ranking UMLS concepts as opposed to documents is better suited for medical professionals who wish to quickly browse diagnoses and their corresponding UMLS concepts, without spending time reading their descriptions. For query 25 seen above, this option identifies the correct UMLS concept \emph{C0268579 Ketotic Hyperglycinemia, Propionic Acidemia, propionyl-CoA carboxylase deficiency, PCC Deficiency} and ranks it as the second most relevant concept to this query (as opposed to rank 4 with standard search). Similarly, for query 18, this option identifies the correct UMLS concept and displays it at rank 5 (as opposed to rank 10 with standard search).

\section{Discussion}
\label{sec:discussion}
One of Google's advantages in web search is its specialized ranking algorithm optimised to work with a large sized index. Our finding, that FindZebra outperforms Google overall for this task and especially when restricted to the sites of our collection (Google Restricted), suggests that Google ranking algorithm is suboptimal for the task at hand. The poor Google Restricted results highlight this because in this case FindZebra and Google are using the same limited, focused data. When broadening the data collection indexed by Google using again a Google Custom Search, but which searches the entire web, only emphasising the documents in the limited collection (Google Custom), the performance of Google is improved but still inferior to the evaluation results obtained for FindZebra. PubMed also has a large index, containing a comprehensive resource of medical articles, however the search approach is different, as results cannot be ranked by relevance but exact match is expected. While this can be overcome by boolean queries, the query complexity becomes quite high and the amount spent on constructing such queries would be more than what is thought of reasonable in medical literature.

It is probably the case that neither PubMed nor Google handle long queries very well, as is the case with long lists of symptoms and observations. This might explain why FindZebra is able to achieve better results for HLJ and OJRD queries, which have an average query length of 28 and 21 words, respectively. For BMJ queries, which have a much lower average query length of 5 words, being devised specifically for Google Search, FindZebra is on equal footing with Google Custom, with Google Search, Google Restricted and PubMed performing considerably better than on the OJRD queries.

In addition to retrieving relevant documents at higher ranks than Google, our system also returns correct results for more queries in the top 20 retrieved documents (67.9\%) than any of the Google variants we tested (37.5\% at best). For the specific task described in this work, we can argue that the most important success consideration is for the correct diagnosis to appear at the top of the diagnostic hypotheses returned by the system. In our case, we not only get in 67.9\% of the cases the correct result in top 20, but it is important to also note that the returned results are actually disease hypotheses, streamlining the clinician's diagnostic process.

The MRR scores show that on average the correct diagnosis appears above rank 3 with our system (0.385), at around rank 5 with Google Search (0.206) and lower for Google Restricted, Google Custom and PubMed. Standard Google Search can thus, to some degree, compensate for the optimal design (in this context) by its larger index. What the MRR actually means for a clinician is that, with FindZebra, when selecting the diagnostic hypotheses, it would be enough to include on average only the first three disease hypotheses retrieved using the system. Moreover, as the results correspond to diseases, transforming a search result into a diagnostic hypothesis should be straightforward. Each result is associated with a description of the disease from a reputable source. Compare this to Google, where results can be retrieved from various types of sources and reading through search results snippets might not be as straightforward.

The evaluation has revealed 5 queries (5, 31, 40, 42, and 52) for which Google Custom returns the correct result and FindZebra not. This result shows that FindZebra, despite being the overall better, still has limitations. The FindZebra index did not include documents for diseases for query 40 and 42, but we have multiple documents for the others. It is notable that 4 of the 5 queries are BMJ queries, which are considerably shorter than both HLJ and OJRD queries. One work-around for short queries would be to enrich the query with synonyms and conceptually similar medical terms. This technique, known as query expansion, is commonly employed in information retrieval systems and also by Google\footnote{\url{http://googleblog.blogspot.com/2010/01/helping-computers-understand-language.html}.} The articles retrieved by Google Custom pointing to the correct diagnosis were in all 5 cases not indexed by FindZebra. This suggests that adding more sources, such as UpToDate and Medline Plus, could improve the system. However, he have yet to assess how the performance is affected by the number of sources used and how each source contributes to the overall quality of the system.

Finally, it is worthwhile discussing how well our evaluation approach can mimic a real diagnostic situation. The main limitation is that it is hard to achieve a query completely blind to the diagnosis because it is known at the time of the construction of the query. The queries are based upon descriptions of the symptoms from case stories. Knowing the diagnosis will probably to some degree bias the description of symptoms. To illustrate how such a bias may occur, the authors of the BMJ queries state that "\ldots although we were blinded to the correct diagnosis, one author was a respiratory and sleep trainee and the other a rheumatologist; sometimes the diagnoses were evident to us, and this could have affected our choice of search terms" \cite{tang2006googling}.

Another aspect is to what degree expert knowledge goes into construction of the queries. From an efficiency point of view we would like the work going into this to be as little as possible. However, the key to successful retrieval is the clinician's ability to represent observations on symptoms in terms that resonate with the general usage as well as with the ranking algorithm used. It is not clear that a typical description of a case appearing in a journal fits web search well. The authors of the BMJ queries are quite clear that medical judgement went into creating their queries: "We chose between three to five search terms for each case, depending on symptoms and signs that we felt would not return a non-specific result. We selected “statistically improbable phrases” whenever possible” \cite{tang2006googling}. When comparing to the original cases taken from the New England Journal of Medicine referenced in \cite{tang2006googling}, it is apparent that the authors did more than just select symptoms from the original reports. They often changed the words, used synonyms, and employed high-level knowledge to arrive at the BMJ queries. For the OJRD queries we used the original descriptions from the article. It could therefore be of interest to follow the same procedure for the BMJ queries and compare performance before and after the adaptation to web search. We suspect that the original longer descriptions of the cases would probably be beneficial for FindZebra but not for Google Search.

\section{Summary and conclusion}
\label{sec:conclusion}
Effective text processing tools are very important to aid biomedical researchers. There has been a remarkable surge of new advances in biomedical language processing, and web search engines in particular are becoming increasingly popular for the task of diagnosing difficult cases. In this article we have asked ourselves how effective is web search actually for diagnosis? We therefore designed an evaluation approach and focussed on the most popular resources used namely Google Search and PubMed. In order to address what determine successes and failures of these resources we developed FindZebra, a specialized search engine for rare diseases, customised for rare disease diagnosis by clinicians in terms of the selection of curated data resources, system interface, indexing (e.g. associating UMLS medical concept identifiers to indexed documents) and retrieval functionalities. The evaluation convincingly showed that FindZebra outperformed Google on standardised performance metrics, specifically precision at rank $k$ (with $k = 10, 20$), mean reciprocal rank and the percentage of queries for which the correct result was returned.

A lesson of this study and similar work in this field is that the ranking algorithm used by large-scale web search engines like Google are not optimized for particular unusual domains, making it feasible to build improved specialized search engines. Hence, one of the contributions of this work is to demonstrate how one can ‘do more with less’: we simply used an open source information retrieval system with standard settings and freely available online medical information. The perspective of combining the ease-of-use of web search with specialized domain knowledge should be attractive to specialists in many areas as it has the potential to greatly improve the quality of search as our work of diagnosis of rare diseases has demonstrated.

There are several ways to move this work forward. On one side we may go further along the path set out in this work by using more advanced information retrieval models and collect data from additional authoritative sources. A perhaps even better strategy would be to work on the data acquisition side and directly and correctly collect symptom-diagnosis association data. In setting up our evaluation approach we found it surprisingly difficult to collect queries with an associated diagnosis. Initiatives in the medical community to systematically collect this kind of data in an unbiased way would be a valuable source for better information retrieval system performance and precision assessment.

\section*{Authors contribution}
R.D., P.P., C.L. and O.W. overall contributed equally to the project. R.D. and P.P. collected the data, set up the search engine, performed the evaluation under the guidance of C.L. and O.W. C.L. and O.W. wrote the paper assisted by R.D. and P.P. The remainder of the authors commented on the paper. O.W. conceived the project and the other authors primarily R.D., P.P. and C.L. contributed to further development of the project.

\section*{Conflicts of interest}
None declared

\section*{Acknowledgements}
The authors wish to thank MD Lennart Friis-Hansen for the initial inspiration for the project.

\newpage



\section*{Supplementary material}

# Appendix: Supplementary material

## Supplementary Table 1 - Queries

This table displays the real-life medical cases of rare diseases used for the experimental evaluation of this work. The sources are HLJ (MD Henrik L. Jørgensen), References [31] to [50] are from the Orphanet Journal of Rare Diseases (OJRD) and BMJ is referring to the British Medical Journal (BMJ) article of Tang and Ng [20].

| No | Query | Diagnosis | Source |
|---|---|---|---|
| 1 | Boy, normal birth, deformity of both big toes (missing joint), quick development of bone tumor near spine and osteogenesis at biopsy | Fibrodysplasia ossificans progressiva | HLJ |
| 2 | Normally developed boy age 5, progressive development of talking difficulties, seizures, ataxia, adrenal insufficiency and degeneration of visual and auditory functions | Adrenoleukodystrophy autosomal neonatal form | HLJ |
| 3 | Boy age 14, yellow, keratotic plaques on the skin of palms and soles going up onto the dorsal side. Both hands and feet are affected | Papillon Lefevre syndrome | HLJ |
| 4 | Jewish boy age 16, monthly seizures, sleep deficiency, aggressive and irritable when woken, highly increased sexual appetite and hunger | Kleine Levin Syndrome | HLJ |
| 5 | Male child, malformations at birth, midfacial retraction with a deep groove under the eyes, and hypertelorism, short nose with a low nasal bridge and large lowset ears, wide mouth and retrognathia, Hypertrichosis with bright reddish hair and a median frontal cutaneous angioma, short neck with redundant skin, Bilateral inguinal hernias, hypospadias with a megameatus, and cryptorchidism | Schinzel-Giedion Syndrome | HLJ |
| 6 | 6 year old, girl, weight length head circumference below the third percentile, atrophic and hyperpigmented skin lesions, pointed nose, aberrant thumbs with diminished flexion, bilateral glue ears, purulent rhinitis | Rothmund-Thomson syndrome | [31] |
| 7 | 13 year old, teenage girl, skeletal muscle defects (muscle weakness), mild mental retardation, ophthalmoparesis | Autosomal recessive centronuclear myopathy (ARCNM) | [32] |
| 8 | 14 year old, teenage boy, mild mental retardation, proximal muscle weakness, unable to walk (wheelchair-bound), premature ventricular complexes, ophthalmoparesis | Autosomal recessive centronuclear myopathy (ARCNM) | [32] |
| 9 | 35 year old, female, progressive disturbance of gait (difficulties in walking), recurrent diarrhea, bronchitis, growth retardation, mild retardation of psychomotor development in infancy, bilateral juvenile cataracts, swelling of the Achilles tendons, high arched feet, exaggerated tendon reflexes | Cerebrotendinous xanthomatosis (CTX) | [33] |
| 10 | 25 year old, woman, conjunctival hyperaemia, interstitial keratitis, moderate bilateral sensorineural hearing loss, tinnitus, dizziness, nausea and vertigo | Cogan's syndrome | [34] |
| 11 | 11 year old, boy, severe psychomotor retardation, seizures, strabismus, inverted nipples, dilated cardiomyopathy, hypotonia, wheelchair-bound | CDG (Congenital Disorders of Glycosylation) syndrome type Ic. *(Synonyms: Carbohydrate deficient glycoprotein syndrome type Ic, Congenital disorder of glycosylation type 1c (or Ic))* | [35] |
| 12 | 17 year old, woman, congenital right pulmonary hypoplasia, right hip dysplasia, absence of uterus, rudimentary uterine horn | Mayer-Rokitansky-Küster-Hauser syndrome | [36] |
| 13 | 10 year old, girl, thrombocytopenia, splenomegaly, headache, itching rubeoliform rash | Congenital hepatic fibrosis (CHF) | [37] |
| 14 | 11 year old, girl, intermittent abdominal pain, mild dorsal scoliosis, low serum phosphate/hypophosphatemia, hypercalciuria, elevated serum 1,25 dihydroxyvitamin D | Hypophosphatemic rickets with hypercalciuria | [38] |
| 15 | 4 month old, boy, epistaxis, haematemesis, haematochezia, subconjunctival bleeding, petechiae, haematomas, haemangioma, slightly enlarged liver, elevated serum transaminases | Type I tyrosinemia. *(Synonyms: Fumarylacetoacetase deficiency, Hepatorenal tryosinosis/tyrosinemia)* | [39] |
| 16 | 7 year old, boy, dysmorphic signs, blue sclerae, high-arched palate, bifid uvula, joint hypermobility, muscular hypotrophy, translucent skin, aortic root dilatation, camptodactyly and ulnar deviation | Loeys-Dietz syndrome (LDS) type I | [40] |
| 17 | 48 year old, woman, aortic aneurysm, haematoma, translucent skin, bilateral venous varicosities, recurrent wrist dislocations | Loeys-Dietz syndrome (LDS) type II | [40] |
| 18 | 8 months old, male, progressive signs of respiratory distress, tachypnea, pulmonary hypertension, tortuosity of aortic arch, facial dysmorphisms | Arterial tortuosity syndrome (ATS) | [41] |
| 19 | 5 year old, male, dyspnoea, asthenia, pulmonary hypertension, severe stenoses elongation and tortuosity of pulmonary arteries branches aortic arch sovraortic trunks and iliac arteries, dysmorphic features, joints hypermobility | Arterial tortuosity syndrome (ATS) | [41] |
| 20 | 64 year old, male, inflammatory back pain, flares of arthritis, multisegmental spondylitis | Whipple's disease. *(Synonyms: Intestinal lipodystrophy, Intestinal lipophagie granulomatosis, Secondary non-tropical sprue)* | [42] |
| 21 | 70 year old, male, massive hemoptysis, respiratory distress, anemia, hemodynamic instability, renal failure, intense headache, arthralgia, myalgias, ecchymoses over | Pulmonary hemorrhage syndrome associated with dengue fever/dengue | [43] |

| | | | |
|---|---|---|---|
| | arms and abdomen, acidosis, pleural effusions, blood tinged secretion from lungs | hemorrhagic fever | |
| 22 | 46 year old, female, ptosis, acanthocytosis, history of diarrhea, ataxia, paresthesia | Abetalipoproteinemia (ABL). *(Synonyms: Bassen-Kornzweig disease, Homozygous familial hypobetalipoproteinemia (HoFHBL))* | [44] |
| 23 | 16 year old, girl, persistent diarrhea, acanthocytosis, mild dysarthria, reduced muscle bulk, bilateral proximal muscle weakness, absent deep-tendon reflexes, upgoing plantar reflexes, reduced sensitivity to light, dysdiadochokinesia | Abetalipoproteinemia (ABL). *(Synonyms: Bassen-Kornzweig disease, Homozygous familial hypobetalipoproteinemia (HoFHBL))* | [44] |
| 24 | teenager, girl, hypotonia, dehydration, acidosis, massive ketonuria, hyperammonemia | Methylmalonic acidemia (MMA). *(Synonyms: Methylmalonic aciduria)* | [45] |
| 25 | girl, hypotonia, seizures, dehydration, polypnea, acidosis, massive ketonuria, hyperammonemia | Propionic acidemia (PA). *(Synonyms: Propionic aciduria, Ketotic glycinemia, Propionyl-CoA carboxylase defficiency )* | [45] |
| 26 | 27 year old, woman, blindness, obesity, type 2 diabetes, renal dysfunction, chronic pyelonephritis, hypertension, hirsutism, retinitis pigmentosa, cataract | Alstrom syndrome (Alström syndrome) | [46] |
| 27 | 17 year old, boy, lysinuric protein intolerance, mild restrictive functional impairment, digital clubbing, atypical abdominal and thoracic pain, ground glass attenuation, interlobular septa thickening, moderate restrictive ventilatory defect, mild anemia, thrombocytopenia, increase in lactate dehydrogenase | Pulmonary alveolar proteinosis (PAP) | [47] |
| 28 | girl, pronounced microcephaly, short stature, psychomotoric delay, distinctive facial appearance, thrombocytopenia, anemia, leukocytopenia, pancytopenia, growth retardation, telecanthus, epicanthal folds, ptosis, infections of the inner ear and respiratory tract, hypoplastic marrow with cellular dysplasia | Ligase IV deficiency syndrome (LIG4 syndrome) *(Synonyms: Ligase 4 syndrome)* | [48] |
| 29 | 5 year old, boy, congenital malformations, malformations of the hands and feet, bilateral strabismus, small tongue, impaired coordination, expressionless face, prominent forehead, depressed nasal bridge, hypoplastic thumbs, bilateral adactyly of the feet, short stature, severe myopia | Oromandibular-limb hypogenesis-Möbius syndrome | [49] |
| 30 | 21 year old, female, irregular menses, menorrhagia, hand and foot malformation, ovarian cyst, basic cognitive function | Terminal deletion of chromosome 4q | [50] |
| 31 | Acute Aortic regurgitation, depression, abscess | Infective endocarditis | BMJ |
| 32 | oesophageal cancer, refractory hic cups, nausea, vomiting | Linitis plastica with bowel obstruction | BMJ |
| 33 | hypertension, adrenal mass | Cushings secondary to adrenal adenoma | BMJ |
| 34 | hip lesion, older child | Osteoid osteoma | BMJ |
| 35 | HRCT centrilobular nodules, acute respiratory failure | Hot tub lung secondary to M avium | BMJ |
| 36 | fever, bilateral thigh pain, weakness | Ehrlichiosis | BMJ |
| 37 | fever, anterior mediastinal mass and central necrosis | Lymphoma | BMJ |
| 38 | multiple spinal tumours, skin tumours | Neurofibromatosis type 1 | BMJ |
| 39 | ulcerative colitis, blurred vision, fever | Vasculitis | BMJ |
| 40 | nephrotic syndrome, Bence Jones, ventricular failure | Amyloid light chain | BMJ |
| 41 | hypertension, papilledema, headache, renal mass, cafe au lait | Pheochromocytoma | BMJ |
| 42 | sickle cell, pulmonary infiltrates, back pain | Acute chest syndrome | BMJ |
| 43 | fibroma, astrocytoma, tumor, leiomyoma, scoliosis | Endometriosis | BMJ |
| 44 | pulmonary infiltrates, cns lesion | Aspiration pneumonia and brain abscess (polymicrobial) | BMJ |
| 45 | CLL, encephalitis | West Nile fever | BMJ |
| 46 | portal vein thrombosis, cancer | Pylephlebitis | BMJ |
| 47 | cardiac arrest, exercise, young | Hypertrophic Obstructive Cardiomyopathy (HOCM) | BMJ |
| 48 | ataxia, confusion, insomnia, death | Creutzfeldt-Jakob disease (CJD) | BMJ |
| 49 | wheeze wt loss, ANCA, haemoptysis, haematuria | Churg Strauss | BMJ |
| 50 | myopathy, neoplasia, dysphagia, rash, periorbital swelling | Dermatomyositis secondary to NHL | BMJ |
| 51 | renal transplant, fever, cat, lymphadenopathy | Cat scratch disease | BMJ |
| 52 | buttock rash, renal failure, edema | Cryoglobulinaemia | BMJ |
| 53 | polyps, telangectasia, epistaxis, anemia | MADH4 mutation (HTT + juvenile polyposis) | BMJ |
| 54 | bullous skin conditions, respiratory failure, carbamazepine | Toxic Epidermal Necrolysis Syndrome | BMJ |

| | | | | | | | | | | | |
|---|---|---|---|---|---|---|---|---|---|---|---|
| | | | | | | (TENS) | | | | | |
| 55 | seizure, confusion, dysphasia, T2 lesions | | | | | | MELAS | | | | BMJ |
| 56 | cardiac arrest sleep | | | | | | Brugada | | | | BMJ |

## Supplementary Table 2 - Evaluation Results (Detailed)

This table breaks down the results of Table 2 for each query. Rank 1st relevant is the rank of the first relevant document (the lower the rank number, the better the performance). Relevant @10/20 is the number of relevant documents in the top 10/20 ranks (the higher the number, the better the performance).

| | FindZebra | | | Google Search | | | Google Custom | | | Google Restricted | | | PubMed | | |
|---|---|---|---|---|---|---|---|---|---|---|---|---|---|---|---|
| | Rank | Relevant | | Rank | Relevant | | Rank | Relevant | | Rank | Relevant | | Rank | Relevant | |
| Query | 1st rel. | @10 | @20 | 1st rel. | @10 | @20 | 1st rel. | @10 | @20 | 1st rel. | @10 | @20 | 1st rel. | @10 | @20 |
| 1 | 1 | 2 | 2 | 10 | 1 | 1 | - | 0 | 0 | - | 0 | 0 | - | 0 | 0 |
| 2 | 2 | 2 | 4 | - | 0 | 0 | 1 | 1 | 2 | 1 | 1 | 1 | - | 0 | 0 |
| 3 | 9 | 1 | 1 | - | 0 | 0 | - | 0 | 0 | - | 0 | 0 | - | 0 | 0 |
| 4 | 1 | 3 | 4 | - | 0 | 0 | 1 | 1 | 1 | - | 0 | 0 | - | 0 | 0 |
| 5 | - | 0 | 0 | 1 | 2 | 2 | 1 | 1 | 1 | - | 0 | 0 | - | 0 | 0 |
| 6 | - | 0 | 0 | - | 0 | 0 | - | 0 | 0 | - | 0 | 0 | - | 0 | 0 |
| 7 | 1 | 2 | 2 | - | 0 | 0 | 17 | 0 | 1 | - | 0 | 0 | - | 0 | 0 |
| 8 | 2 | 2 | 2 | - | 0 | 0 | - | 0 | 0 | - | 0 | 0 | - | 0 | 0 |
| 9 | 6 | 1 | 1 | - | 0 | 0 | - | 0 | 0 | - | 0 | 0 | - | 0 | 0 |
| 10 | 7 | 1 | 1 | 7 | 1 | 2 | - | 0 | 0 | - | 0 | 0 | - | 0 | 0 |
| 11 | 1 | 2 | 3 | 3 | 2 | 2 | - | 0 | 0 | - | 0 | 0 | - | 0 | 0 |
| 12 | 5 | 2 | 2 | 10 | 1 | 1 | - | 0 | 0 | - | 0 | 0 | - | 0 | 0 |
| 13 | - | 0 | 0 | - | 0 | 0 | - | 0 | 0 | - | 0 | 0 | - | 0 | 0 |
| 14 | 1 | 4 | 6 | - | 0 | 0 | - | 0 | 0 | - | 0 | 0 | - | 0 | 0 |
| 15 | - | 0 | 0 | - | 0 | 0 | - | 0 | 0 | - | 0 | 0 | - | 0 | 0 |
| 16 | 1 | 4 | 5 | - | 0 | 0 | - | 0 | 0 | - | 0 | 0 | - | 0 | 0 |
| 17 | 5 | 2 | 4 | - | 0 | 0 | - | 0 | 0 | - | 0 | 0 | - | 0 | 0 |
| 18 | 10 | 1 | 1 | - | 0 | 0 | - | 0 | 0 | - | 0 | 0 | - | 0 | 0 |
| 19 | 1 | 2 | 3 | - | 0 | 0 | - | 0 | 0 | - | 0 | 0 | - | 0 | 0 |
| 20 | - | 0 | 0 | - | 0 | 0 | - | 0 | 0 | - | 0 | 0 | - | 0 | 0 |
| 21 | - | 0 | 0 | - | 0 | 0 | - | 0 | 0 | - | 0 | 0 | - | 0 | 0 |
| 22 | 1 | 5 | 6 | - | 0 | 0 | - | 0 | 0 | - | 0 | 0 | - | 0 | 0 |
| 23 | - | 0 | 0 | - | 0 | 0 | - | 0 | 0 | - | 0 | 0 | - | 0 | 0 |
| 24 | 1 | 4 | 5 | - | 0 | 0 | - | 0 | 0 | - | 0 | 0 | - | 0 | 0 |
| 25 | 4 | 2 | 2 | - | 0 | 0 | - | 0 | 0 | - | 0 | 0 | - | 0 | 0 |
| 26 | 2 | 2 | 2 | - | 0 | 0 | - | 0 | 0 | - | 0 | 0 | - | 0 | 0 |
| 27 | 1 | 2 | 3 | - | 0 | 0 | - | 0 | 0 | - | 0 | 0 | - | 0 | 0 |
| 28 | - | 0 | 0 | - | 0 | 0 | - | 0 | 0 | - | 0 | 0 | - | 0 | 0 |
| 29 | 1 | 1 | 1 | - | 0 | 0 | - | 0 | 0 | - | 0 | 0 | - | 0 | 0 |
| 30 | - | 0 | 0 | - | 0 | 0 | - | 0 | 0 | - | 0 | 0 | - | 0 | 0 |
| 31 | - | 0 | 0 | - | 0 | 0 | 4 | 4 | 6 | - | 0 | 0 | 1 | 1 | 1 |
| 32 | - | 0 | 0 | - | 0 | 0 | - | 0 | 0 | - | 0 | 0 | - | 0 | 0 |
| 33 | 10 | 1 | 2 | 1 | 1 | 1 | 7 | 1 | 2 | - | 0 | 0 | 1 | 3 | 3 |
| 34 | 19 | 0 | 1 | - | 0 | 0 | 9 | 1 | 1 | - | 0 | 0 | - | 0 | 0 |
| 35 | 7 | 1 | 1 | - | 0 | 0 | 3 | 2 | 2 | 1 | 1 | 1 | - | 0 | 0 |
| 36 | - | 0 | 0 | - | 0 | 0 | - | 0 | 0 | - | 0 | 0 | - | 0 | 0 |
| 37 | 7 | 1 | 1 | 1 | 2 | 2 | 4 | 1 | 4 | - | 0 | 0 | - | 0 | 0 |
| 38 | 1 | 1 | 3 | 4 | 2 | 4 | 1 | 3 | 3 | 1 | 2 | 2 | 11 | 0 | 3 |
| 39 | 8 | 1 | 1 | - | 0 | 0 | - | 0 | 0 | - | 0 | 0 | - | 0 | 0 |
| 40 | - | 0 | 0 | 1 | 5 | 10 | 1 | 5 | 10 | - | 0 | 0 | - | 0 | 0 |
| 41 | 13 | 0 | 1 | 13 | 0 | 1 | 5 | 2 | 2 | - | 0 | 0 | - | 0 | 0 |
| 42 | - | 0 | 0 | 2 | 5 | 8 | 4 | 2 | 5 | - | 0 | 0 | 1 | 1 | 1 |
| 43 | - | 0 | 0 | - | 0 | 0 | - | 0 | 0 | - | 0 | 0 | - | 0 | 0 |
| 44 | - | 0 | 0 | - | 0 | 0 | - | 0 | 0 | - | 0 | 0 | - | 0 | 0 |
| 45 | 17 | 0 | 2 | 18 | 0 | 1 | 15 | 0 | 1 | - | 0 | 0 | - | 0 | 0 |
| 46 | - | 0 | 0 | - | 0 | 0 | - | 0 | 0 | - | 0 | 0 | - | 0 | 0 |
| 47 | 5 | 2 | 3 | 1 | 1 | 1 | 12 | 0 | 1 | - | 0 | 0 | 12 | 0 | 3 |
| 48 | 1 | 2 | 4 | - | 0 | 0 | - | 0 | 0 | - | 0 | 0 | - | 0 | 0 |
| 49 | 3 | 1 | 1 | 1 | 2 | 2 | 16 | 0 | 1 | - | 0 | 0 | - | 0 | 0 |
| 50 | 1 | 2 | 3 | 1 | 6 | 12 | 1 | 8 | 11 | 1 | 1 | 1 | 1 | 1 | 1 |
| 51 | 1 | 2 | 4 | 1 | 5 | 7 | 1 | 8 | 12 | - | 0 | 0 | 1 | 3 | 3 |
| 52 | - | 0 | 0 | - | 0 | 0 | 1 | 2 | 2 | - | 0 | 0 | - | 0 | 0 |
| 53 | 1 | 3 | 5 | - | 0 | 0 | 2 | 3 | 4 | - | 0 | 0 | 1 | 2 | 2 |

| 54 | 1 | 3 | 4 | 1 | 2 | 5 | 1 | 3 | 4 | 1 | 1 | 1 | 1 | 1 | 1 |
| 55 | 3 | 1 | 1 | - | 0 | 0 | - | 0 | 0 | - | 0 | 0 | - | 0 | 0 |
| 56 | 3 | 2 | 3 | 1 | 1 | 1 | - | 0 | 0 | 2 | 1 | 1 | - | 0 | 0 |

*Supplementary Table 3 - queries for which correct/relevant results were returned (detailed)*

This table breaks down the results for the queries by system. The following is a list of comments about specific difficult queries. HLJ on query 13: These symptoms could be caused by many different diseases including some fairly common ones. It might not be the best case story for your purposes. On the other hand, I think that it will be interesting to see what the system will come up with here. HLJ on query 20: This case is interesting but I doubt the system will work here. In a patient of 64 years, these symptoms could be caused by a multitude of diseases, most of them much more common than the rare infectious disease. HLJ on query 21: This one is also interesting, although not that uncommon. Several other similar infections could produce a picture like this, so I am looking forward to the proposals of the system. BMJ authors on query 36: less likely to be successful because of complex disease with non-specific symptoms. BMJ authors on query 43: less likely to be successful because of common disease with rare presentation.

| Query Source | Query No | Search Terms | Final diagnosis | Systems with relevant results | FindZebra | Google Search | Google Custom | Google Restricted | Pubmed |
|---|---|---|---|---|---|---|---|---|---|
| HLJ | 1 | Boy, normal birth, deformity of both big toes (missing joint), quick development of bone tumor near spine and osteogenesis at biopsy | Fibrodysplasia ossificans progressiva | 2 | 1 | 1 | 0 | 0 | 0 |
| HLJ | 2 | Normally developed boy age 5, progressive development of talking difficulties, seizures, ataxia, adrenal insufficiency and degeneration of visual and auditory functions | Adrenoleukodystrophy autosomal neonatal form | 3 | 1 | 0 | 1 | 1 | 0 |
| HLJ | 3 | Boy age 14, yellow, keratotic plaques on the skin of palms and soles going up onto the dorsal side. Both hands and feet are affected | Papillon Lefevre syndrome | 1 | 1 | 0 | 0 | 0 | 0 |
| HLJ | 4 | Jewish boy age 16, monthly seizures, sleep deficiency, aggressive and irritable when woken, highly increased sexual appetite and hunger | Kleine Levin Syndrome | 2 | 1 | 0 | 1 | 0 | 0 |
| HLJ | 5 | Male child, malformations at birth, midfacial retraction with a deep groove under the eyes, and hypertelorism, short nose with a low nasal bridge and large lowset ears, wide mouth and retrognathia. Hypertrichosis with bright reddish hair and a median frontal cutaneous angioma, short neck with redundant skin, Bilateral inguinal hernias, hypospadias with a megameatus, and cryptorchidism | Schinzel-Giedion Syndrome | 2 | 0 | 1 | 1 | 0 | 0 |
| OJRD | 6 | 6 year old, girl, weight length head circumference below the third percentile, atrophic and hyperpigmented skin lesions, pointed nose, aberrant thumbs with diminished flexion, bilateral glue ears, purulent rhinitis | Rothmund-Thomson syndrome | 0 | 0 | 0 | 0 | 0 | 0 |
| OJRD | 7 | 13 year old, teenage girl, skeletal muscle defects (muscle weakness), mild mental retardation, ophthalmoparesis | Autosomal recessive centronuclear myopathy (ARCNM) | 2 | 1 | 0 | 1 | 0 | 0 |
| OJRD | 8 | 14 year old, teenage boy, mild mental retardation, proximal muscle weakness, unable to walk (wheelchair-bound), premature ventricular complexes, ophthalmoparesis | Autosomal recessive centronuclear myopathy (ARCNM) | 1 | 1 | 0 | 0 | 0 | 0 |
| OJRD | 9 | 35 year old, female, progressive disturbance of gait (difficulties in walking), recurrent diarrhea, bronchitis, growth retardation, mild retardation of psychomotor development in infancy, bilateral juvenile cataracts, swelling of the Achilles tendons, high arched feet, exaggerated tendon reflexes | Cerebrotendinous xanthomatosis (CTX) Synonym: Sterol 27-hydroxylase deficiency | 1 | 1 | 0 | 0 | 0 | 0 |
| OJRD | 10 | 25 year old, woman, conjunctival hyperaemia, interstitial keratitis, moderate bilateral sensorineural hearing loss, tinnitus, dizziness, nausea and vertigo | Cogan's syndrome | 2 | 1 | 1 | 0 | 0 | 0 |

| | | | | | | | | | |
|---|---|---|---|---|---|---|---|---|---|
| OJRD | 11 | 11 year old, boy, severe psychomotor retardation, seizures, strabismus, inverted nipples, dilated cardiomyopathy, hypotonia, wheelchair-bound | CDG (Congenital Disorders of Glycosylation) syndrome type Ic Synonyms: Carbohydrate deficient glycoprotein syndrome type Ic, Congenital disorder of glycosylation type 1c (or Ic) | 2 | 1 | 1 | 0 | 0 | 0 |
| OJRD | 12 | 17 year old, woman, congenital right pulmonary hypoplasia, right hip dysplasia, absence of uterus, rudimentary uterine horn | Mayer-Rokitansky-Küster-Hauser syndrome | 2 | 1 | 1 | 0 | 0 | 0 |
| OJRD | 13 | 10 year old, girl, thrombocytopenia, splenomegaly, headache, itching rubeoliform rash | Congenital hepatic fibrosis (CHF) | 0 | 0 | 0 | 0 | 0 | 0 |
| OJRD | 14 | 11 year old, girl, intermittent abdominal pain, mild dorsal scoliosis, low serum phosphate / hypophosphatemia, hypercalciuria, elevated serum 1.25 dihydroxyvitamin D | Hypophosphatemic rickets with hypercalciuria | 1 | 1 | 0 | 0 | 0 | 0 |
| OJRD | 15 | 4 month old, boy, epistaxis, haematemesis, haematochezia, subconjunctival bleeding, petechiae, haematomas, haemangioma, slightly enlarged liver, elevated serum transaminases | Type I tyrosinemia Synonyms: Fumarylacetoacetase deficiency, Hepatorenal tryosinosis / tyrosinemia | 0 | 0 | 0 | 0 | 0 | 0 |
| OJRD | 16 | 7 year old, boy, dysmorphic signs, blue sclerae, high-arched palate, bifid uvula, joint hypermobility, muscular hypotrophy, translucent skin, aortic root dilatation, camptodactyly and ulnar deviation | Loeys-Dietz syndrome (LDS) type I | 1 | 1 | 0 | 0 | 0 | 0 |
| OJRD | 17 | 48 year old, woman, aortic aneurysm, haematoma, translucent skin, bilateral venous varicosities, recurrent wrist dislocations | Loeys-Dietz syndrome (LDS) type II | 1 | 1 | 0 | 0 | 0 | 0 |
| OJRD | 18 | 8 months old, male, progressive signs of respiratory distress, tachypnea, pulmonary hypertension, tortuosity of aortic arch, facial dysmorphisms | Arterial tortuosity syndrome (ATS) | 1 | 1 | 0 | 0 | 0 | 0 |
| OJRD | 19 | 5 year old, male, dyspnoea, asthenia, pulmonary hypertension, severe stenoses elongation and tortuosity of pulmonary arteries branches aortic arch sovraortic trunks and iliac arteries, dysmorphic features, joints hypermobility | Arterial tortuosity syndrome (ATS) | 1 | 1 | 0 | 0 | 0 | 0 |
| OJRD | 20 | 64 year old, male, inflammatory back pain, flares of arthritis, multisegmental spondylitis | Whipple's disease Synonyms: Intestinal lipodystrophy, Intestinal lipophagic granulomatosis, Secondary non-tropical sprue | 0 | 0 | 0 | 0 | 0 | 0 |
| OJRD | 21 | 70 year old, male, massive hemoptysis, respiratory distress, anemia, hemodynamic instability, renal failure, intense headache, arthralgia, myalgias, ecchymoses over arms and abdomen, acidosis, pleural effusions, blood tinged secretion from lungs | Pulmonary hemorrhage syndrome associated with dengue fever /dengue hemorrhagic fever | 0 | 0 | 0 | 0 | 0 | 0 |
| OJRD | 22 | 46 year old, female, ptosis, acanthocytosis, history of diarrhea, ataxia, paresthesia | Abetalipoproteinemia (ABL) Synonyms: Bassen-Kornzweig disease, Homozygous familial hypobetalipoproteinemia (HoFHBL) | 1 | 1 | 0 | 0 | 0 | 0 |
| OJRD | 23 | 16 year old, girl, persistent diarrhea, acanthocytosis, mild dysarthria, reduced muscle bulk, bilateral proximal muscle weakness, absent deep-tendon reflexes, upgoing plantar reflexes, reduced sensitivity to light, dysdiadochokinesia | Abetalipoproteinemia (ABL) Synonyms: Bassen-Kornzweig disease, Homozygous familial hypobetalipoproteinemia (HoFHBL) | 0 | 0 | 0 | 0 | 0 | 0 |
| OJRD | 24 | teenager, girl, hypotonia, dehydration, | Methylmalonic acidemia | 1 | 1 | 0 | 0 | 0 | 0 |

| | | | | | | | | | |
|---|---|---|---|---|---|---|---|---|---|
| | | acidosis, massive ketonuria, hyperammonemia | (MMA) Synonyms: Methylmalonic aciduria | | | | | | |
| OJRD | 25 | girl, hypotonia, seizures, dehydration, polypnea, acidosis, massive ketonuria, hyperammonemia | Propionic acidemia (PA) Synonyms: Propionic aciduria, Ketotic glycinemia, Propionyl-CoA carboxylase deficiency | 1 | 1 | 0 | 0 | 0 | 0 |
| OJRD | 26 | 27 year old, woman, blindness, obesity, type 2 diabetes, renal dysfunction, chronic pyelonephritis, hypertension, hirsutism, retinitis pigmentosa, cataract | Alstrom syndrome (Alström syndrome) | 1 | 1 | 0 | 0 | 0 | 0 |
| OJRD | 27 | 17 year old, boy, lysinuric protein intolerance, mild restrictive functional impairment, digital clubbing, atypical abdominal and thoracic pain, ground glass attenuation, interlobular septa thickening, moderate restrictive ventilatory defect, mild anemia, thrombocytopenia, increase in lactate dehydrogenase | Pulmonary alveolar proteinosis (PAP) | 1 | 1 | 0 | 0 | 0 | 0 |
| OJRD | 28 | girl, pronounced microcephaly, short stature, psychomotoric delay, distinctive facial appearance, thrombocytopenia, anemia, leukocytopenia, pancytopenia, growth retardation, telecanthus, epicanthal folds, ptosis, infections of the inner ear and respiratory tract, hypoplastic marrow with cellular dysplasia | Ligase IV deficiency syndrome (LIG4 syndrome) Synonyms: Ligase 4 syndrome | 0 | 0 | 0 | 0 | 0 | 0 |
| OJRD | 29 | 5 year old, boy, congenital malformations, malformations of the hands and feet, bilateral strabismus, small tongue, impaired coordination, expressionless face, prominent forehead, depressed nasal bridge, hypoplastic thumbs, bilateral adactyly of the feet, short stature, severe myopia | Oromandibular-limb hypogenesis-Möbius syndrome | 1 | 1 | 0 | 0 | 0 | 0 |
| OJRD | 30 | 21 year old, female, irregular menses, menorrhagia, hand and foot malformation, ovarian cyst, basic cognitive function | Terminal deletion of chromosome 4q | 0 | 0 | 0 | 0 | 0 | 0 |
| BMJ | 31 | Acute Aortic regurgitation, depression, abscess | Infective endocarditis | 2 | 0 | 0 | 1 | 0 | 1 |
| BMJ | 32 | oesophageal cancer, refractory hic cups, nausea, vomiting | Linitis plastica with bowel obstruction | 0 | 0 | 0 | 0 | 0 | 0 |
| BMJ | 33 | hypertension, adrenal mass | Cushings secondary to adrenal adenoma | 4 | 1 | 1 | 1 | 0 | 1 |
| BMJ | 34 | hip lesion, older child | Osteoid osteoma | 2 | 1 | 0 | 1 | 0 | 0 |
| BMJ | 35 | HRCT centrilobular nodules, acute respiratory failure | Hot tub lung secondary to M avium | 3 | 1 | 0 | 1 | 1 | 0 |
| BMJ | 36 | fever, bilateral thigh pain, weakness | Ehrlichiosis | 0 | 0 | 0 | 0 | 0 | 0 |
| BMJ | 37 | fever, anterior mediastinal mass and central necrosis | Lymphoma | 3 | 1 | 1 | 1 | 0 | 0 |
| BMJ | 38 | multiple spinal tumours, skin tumours | Neurofibromatosis type 1 | 5 | 1 | 1 | 1 | 1 | 1 |
| BMJ | 39 | ulcerative colitis, blurred vision, fever | Vasculitis | 1 | 1 | 0 | 0 | 0 | 0 |
| BMJ | 40 | nephrotic syndrome, Bence Jones, ventricular failure | Amyloid light chain | 2 | 0 | 1 | 1 | 0 | 0 |
| BMJ | 41 | hypertension, papilledema, headache, renal mass, café au lait | Pheochromocytoma | 3 | 1 | 1 | 1 | 0 | 0 |
| BMJ | 42 | sickle cell, pulmonary infiltrates, back pain | Acute chest syndrome | 3 | 0 | 1 | 1 | 0 | 1 |
| BMJ | 43 | fibroma, astrocytoma, tumor, leiomyoma, scoliosis | Endometriosis | 0 | 0 | 0 | 0 | 0 | 0 |
| BMJ | 44 | pulmonary infiltrates, cns lesion | Aspiration pneumonia and brain abscess (polymicrobial) | 0 | 0 | 0 | 0 | 0 | 0 |
| BMJ | 45 | CLL, encephalitis | West Nile fever | 3 | 1 | 1 | 1 | 0 | 0 |
| BMJ | 46 | portal vein thrombosis, cancer | Pylephlebitis | 0 | 0 | 0 | 0 | 0 | 0 |
| BMJ | 47 | cardiac arrest, exercise, young | Hypertrophic Obstructive Cardiomyopathy (HOCM) | 4 | 1 | 1 | 1 | 0 | 1 |
| BMJ | 48 | ataxia, confusion, insomnia, death | Creutzfeldt-Jakob disease (CJD) | 1 | 1 | 0 | 0 | 0 | 0 |
| BMJ | 49 | wheeze wt loss, ANCA, haemoptysis, haematuria | Churg Strauss | 3 | 1 | 1 | 1 | 0 | 0 |
| BMJ | 50 | myopathy, neoplasia, dysphagia, rash, | Dermatomyositis | 5 | 1 | 1 | 1 | 1 | 1 |

| | | | | | | | | | |
|---|---|---|---|---|---|---|---|---|---|
| | | periorbital swelling | secondary to NHL | | | | | | |
| BMJ | 51 | renal transplant, fever, cat, lymphadenopathy | Cat scratch disease | 4 | 1 | 1 | 1 | 0 | 1 |
| BMJ | 52 | buttock rash, renal failure, edema | Cryoglobulinaemia | 1 | 0 | 0 | 1 | 0 | 0 |
| BMJ | 53 | polyps, telangectasia, epistaxis, anemia | MADH4 mutation (HTT + juvenile polyposis) | 3 | 1 | 0 | 1 | 0 | 1 |
| BMJ | 54 | bullous skin conditions, respiratory failure, carbamazepine | Toxic Epidermal Necrolysis Syndrome (TENS) | 5 | 1 | 1 | 1 | 1 | 1 |
| BMJ | 55 | seizure, confusion, dysphasia, T2 lesions | MELAS | 1 | 1 | 0 | 0 | 0 | 0 |
| BMJ | 56 | cardiac arrest sleep | Brugada | 3 | 1 | 1 | 0 | 1 | 0 |



\biboptions{compress}
\bibliographystyle{model3-num-names}
\bibliography{FindZebra}

\begin{thebibliography}{50}
\providecommand{\natexlab}[1]{#1}
\providecommand{\url}[1]{\texttt{#1}}
\providecommand{\href}[2]{#2}
\providecommand{\path}[1]{#1}
\providecommand{\eprint}[1]{\href{http://arxiv.org/abs/#1}{\path{#1}}}
\providecommand{\DOIprefix}{doi:}
\providecommand{\ArXivprefix}{arXiv:}
\providecommand{\URLprefix}{URL: }
\providecommand{\Pubmedprefix}{pmid:}
\providecommand{\doi}[1]{\href{http://dx.doi.org/#1}{\path{#1}}}
\providecommand{\Pubmed}[1]{\href{pmid:#1}{\path{#1}}}
\providecommand{\BIBand}{and}
\providecommand{\bibinfo}[2]{#2}
\ifx\xfnm\undefined \def\xfnm[#1]{\unskip,\space#1}\fi
\bibitem[{Cartright et~al.(2011)Cartright, White and
  Horvitz}]{cartright2011intentions}
\bibinfo{author}{Cartright\xfnm[ M.]}, \bibinfo{author}{White\xfnm[ R.W.]},
  \bibinfo{author}{Horvitz\xfnm[ E.]}.
\newblock \bibinfo{title}{Intentions and attention in exploratory health
  search}.
\newblock In: \bibinfo{booktitle}{Proc ACM SIGIR}. \bibinfo{year}{2011}, p.
  \bibinfo{pages}{65--74}.
\bibitem[{Hunter and Cohen(2006)}]{hunter2006biomedical}
\bibinfo{author}{Hunter\xfnm[ L.]}, \bibinfo{author}{Cohen\xfnm[ K.B.]}.
\newblock \bibinfo{title}{Biomedical language processing: What's beyond
  pubmed?}
\newblock \bibinfo{journal}{Molecular cell}
  \bibinfo{year}{2006};\bibinfo{volume}{21}(\bibinfo{number}{5}):\bibinfo{pages}{589--594}.
\bibitem[{Purcell et~al.(2010)Purcell, Rainie, Mitchell, Rosenstiel and
  Olmstead}]{purcell2010understanding}
\bibinfo{author}{Purcell\xfnm[ K.]}, \bibinfo{author}{Rainie\xfnm[ L.]},
  \bibinfo{author}{Mitchell\xfnm[ A.]}, \bibinfo{author}{Rosenstiel\xfnm[ T.]},
  \bibinfo{author}{Olmstead\xfnm[ K.]}.
\newblock \bibinfo{title}{Understanding the participatory news consumer}.
\newblock \bibinfo{journal}{Pew Internet and American Life Project}
  \bibinfo{year}{2010};\bibinfo{volume}{1}.
\bibitem[{Spink et~al.(2004)Spink, Yang, Jansen, Nykanen, Lorence, Ozmutlu
  et~al.}]{spink2004study}
\bibinfo{author}{Spink\xfnm[ A.]}, \bibinfo{author}{Yang\xfnm[ Y.]},
  \bibinfo{author}{Jansen\xfnm[ J.]}, \bibinfo{author}{Nykanen\xfnm[ P.]},
  \bibinfo{author}{Lorence\xfnm[ D.P.]}, \bibinfo{author}{Ozmutlu\xfnm[ S.]},
  et~al.
\newblock \bibinfo{title}{A study of medical and health queries to web search
  engines}.
\newblock \bibinfo{journal}{Health Information \& Libraries Journal}
  \bibinfo{year}{2004};\bibinfo{volume}{21}(\bibinfo{number}{1}):\bibinfo{pages}{44--51}.
\bibitem[{White and Horvitz(2009)}]{white2009cyberchondria}
\bibinfo{author}{White\xfnm[ R.W.]}, \bibinfo{author}{Horvitz\xfnm[ E.]}.
\newblock \bibinfo{title}{Cyberchondria: studies of the escalation of medical
  concerns in web search}.
\newblock \bibinfo{journal}{ACM Transactions on Information Systems (TOIS)}
  \bibinfo{year}{2009};\bibinfo{volume}{27}(\bibinfo{number}{4}):\bibinfo{pages}{23}.
\bibitem[{Bouwman et~al.(2010)Bouwman, Teunissen, Wijburg and
  Linthorst}]{bouwman2010doctor}
\bibinfo{author}{Bouwman\xfnm[ M.G.]}, \bibinfo{author}{Teunissen\xfnm[ Q.G.]},
  \bibinfo{author}{Wijburg\xfnm[ F.A.]}, \bibinfo{author}{Linthorst\xfnm[
  G.E.]}.
\newblock \bibinfo{title}{‘doctor google’ending the diagnostic odyssey in
  lysosomal storage disorders: parents using internet search engines as an
  efficient diagnostic strategy in rare diseases}.
\newblock \bibinfo{journal}{Archives of disease in childhood}
  \bibinfo{year}{2010};\bibinfo{volume}{95}(\bibinfo{number}{8}):\bibinfo{pages}{642--644}.
\bibitem[{Manning et~al.(2008)Manning, Raghavan and
  Sch{\"u}tze}]{manning2008introduction}
\bibinfo{author}{Manning\xfnm[ C.D.]}, \bibinfo{author}{Raghavan\xfnm[ P.]},
  \bibinfo{author}{Sch{\"u}tze\xfnm[ H.]}.
\newblock \bibinfo{title}{Introduction to information retrieval};
  vol.~\bibinfo{volume}{1}.
\newblock \bibinfo{publisher}{Cambridge University Press Cambridge};
  \bibinfo{year}{2008}.
\bibitem[{Eysenbach and K{\"o}hler(2002)}]{eysenbach2002consumers}
\bibinfo{author}{Eysenbach\xfnm[ G.]}, \bibinfo{author}{K{\"o}hler\xfnm[ C.]}.
\newblock \bibinfo{title}{How do consumers search for and appraise health
  information on the world wide web? qualitative study using focus groups,
  usability tests, and in-depth interviews}.
\newblock \bibinfo{journal}{Bmj}
  \bibinfo{year}{2002};\bibinfo{volume}{324}(\bibinfo{number}{7337}):\bibinfo{pages}{573--577}.
\bibitem[{Lewis(2006)}]{lewis2006seeking}
\bibinfo{author}{Lewis\xfnm[ T.]}.
\newblock \bibinfo{title}{Seeking health information on the internet: lifestyle
  choice or bad attack of cyberchondria?}
\newblock \bibinfo{journal}{Media, Culture \& Society}
  \bibinfo{year}{2006};\bibinfo{volume}{28}(\bibinfo{number}{4}):\bibinfo{pages}{521--539}.
\bibitem[{Hughes et~al.(2009)Hughes, Joshi, Lemonde and
  Wareham}]{hughes2009junior}
\bibinfo{author}{Hughes\xfnm[ B.]}, \bibinfo{author}{Joshi\xfnm[ I.]},
  \bibinfo{author}{Lemonde\xfnm[ H.]}, \bibinfo{author}{Wareham\xfnm[ J.]}.
\newblock \bibinfo{title}{Junior physician's use of web 2.0 for information
  seeking and medical education: a qualitative study}.
\newblock \bibinfo{journal}{International journal of medical informatics}
  \bibinfo{year}{2009};\bibinfo{volume}{78}(\bibinfo{number}{10}):\bibinfo{pages}{645--655}.
\bibitem[{Schwarz and Morris(2011)}]{schwarz2011augmenting}
\bibinfo{author}{Schwarz\xfnm[ J.]}, \bibinfo{author}{Morris\xfnm[ M.]}.
\newblock \bibinfo{title}{Augmenting web pages and search results to support
  credibility assessment}.
\newblock In: \bibinfo{booktitle}{Proceedings of the 2011 annual conference on
  Human factors in computing systems}. \bibinfo{organization}{ACM};
  \bibinfo{year}{2011}, p. \bibinfo{pages}{1245--1254}.
\bibitem[{Hersh and Hickam(1998)}]{hersh1998well}
\bibinfo{author}{Hersh\xfnm[ W.R.]}, \bibinfo{author}{Hickam\xfnm[ D.H.]}.
\newblock \bibinfo{title}{How well do physicians use electronic information
  retrieval systems?}
\newblock \bibinfo{journal}{JAMA: the journal of the American Medical
  Association}
  \bibinfo{year}{1998};\bibinfo{volume}{280}(\bibinfo{number}{15}):\bibinfo{pages}{1347--1352}.
\bibitem[{Bhavnani(2002)}]{bhavnani2002domain}
\bibinfo{author}{Bhavnani\xfnm[ S.K.]}.
\newblock \bibinfo{title}{Domain-specific search strategies for the effective
  retrieval of healthcare and shopping information}.
\newblock In: \bibinfo{booktitle}{CHI'02 extended abstracts on Human factors in
  computing systems}. \bibinfo{organization}{ACM}; \bibinfo{year}{2002}, p.
  \bibinfo{pages}{610--611}.
\bibitem[{Barnett et~al.(1987)Barnett, Cimino, Hupp and
  Hoffer}]{barnett1987evolving}
\bibinfo{author}{Barnett\xfnm[ G.O.]}, \bibinfo{author}{Cimino\xfnm[ J.J.]},
  \bibinfo{author}{Hupp\xfnm[ J.A.]}, \bibinfo{author}{Hoffer\xfnm[ E.P.]}.
\newblock \bibinfo{title}{An evolving diagnostic decision-support system}.
\newblock \bibinfo{journal}{Jama}
  \bibinfo{year}{1987};\bibinfo{volume}{258}:\bibinfo{pages}{67--74}.
\bibitem[{Fryer(1991)}]{fryer1991london}
\bibinfo{author}{Fryer\xfnm[ A.]}.
\newblock \bibinfo{title}{London dysmorphology database}.
\newblock \bibinfo{journal}{Journal of Medical Genetics}
  \bibinfo{year}{1991};\bibinfo{volume}{28}(\bibinfo{number}{10}):\bibinfo{pages}{727}.
\bibitem[{Ten{\'o}rio et~al.(2011)Ten{\'o}rio, Hummel, Cohrs, Sdepanian, Pisa
  and de~F{\'a}tima~Marin}]{tenorio2011artificial}
\bibinfo{author}{Ten{\'o}rio\xfnm[ J.M.]}, \bibinfo{author}{Hummel\xfnm[
  A.D.]}, \bibinfo{author}{Cohrs\xfnm[ F.M.]}, \bibinfo{author}{Sdepanian\xfnm[
  V.L.]}, \bibinfo{author}{Pisa\xfnm[ I.T.]},
  \bibinfo{author}{de~F{\'a}tima~Marin\xfnm[ H.]}.
\newblock \bibinfo{title}{Artificial intelligence techniques applied to the
  development of a decision--support system for diagnosing celiac disease}.
\newblock \bibinfo{journal}{International journal of medical informatics}
  \bibinfo{year}{2011};\bibinfo{volume}{80}(\bibinfo{number}{11}):\bibinfo{pages}{793--802}.
\bibitem[{Ahmadian et~al.(2011)Ahmadian, van Engen-Verheul, Bakhshi-Raiez,
  Peek, Cornet and de~Keizer}]{ahmadian2011role}
\bibinfo{author}{Ahmadian\xfnm[ L.]}, \bibinfo{author}{van Engen-Verheul\xfnm[
  M.]}, \bibinfo{author}{Bakhshi-Raiez\xfnm[ F.]}, \bibinfo{author}{Peek\xfnm[
  N.]}, \bibinfo{author}{Cornet\xfnm[ R.]}, \bibinfo{author}{de~Keizer\xfnm[
  N.F.]}.
\newblock \bibinfo{title}{The role of standardized data and terminological
  systems in computerized clinical decision support systems: Literature review
  and survey}.
\newblock \bibinfo{journal}{International journal of medical informatics}
  \bibinfo{year}{2011};\bibinfo{volume}{80}(\bibinfo{number}{2}):\bibinfo{pages}{81--93}.
\bibitem[{Hersh et~al.(2002)Hersh, Crabtree, Hickam, Sacherek, Friedman,
  Tidmarsh et~al.}]{hersh2002factors}
\bibinfo{author}{Hersh\xfnm[ W.R.]}, \bibinfo{author}{Crabtree\xfnm[ M.K.]},
  \bibinfo{author}{Hickam\xfnm[ D.H.]}, \bibinfo{author}{Sacherek\xfnm[ L.]},
  \bibinfo{author}{Friedman\xfnm[ C.P.]}, \bibinfo{author}{Tidmarsh\xfnm[ P.]},
  et~al.
\newblock \bibinfo{title}{Factors associated with success in searching medline
  and applying evidence to answer clinical questions}.
\newblock \bibinfo{journal}{Journal of the American Medical Informatics
  Association}
  \bibinfo{year}{2002};\bibinfo{volume}{9}(\bibinfo{number}{3}):\bibinfo{pages}{283--293}.
\bibitem[{Lombardi et~al.(2009)Lombardi, Griffiths, McLeod, Caviglia and
  Penagos}]{lombardi2009search}
\bibinfo{author}{Lombardi\xfnm[ C.]}, \bibinfo{author}{Griffiths\xfnm[ E.]},
  \bibinfo{author}{McLeod\xfnm[ B.]}, \bibinfo{author}{Caviglia\xfnm[ A.]},
  \bibinfo{author}{Penagos\xfnm[ M.]}.
\newblock \bibinfo{title}{Search engine as a diagnostic tool in difficult
  immunological and allergologic cases: is google useful?}
\newblock \bibinfo{journal}{Internal medicine journal}
  \bibinfo{year}{2009};\bibinfo{volume}{39}(\bibinfo{number}{7}):\bibinfo{pages}{459--464}.
\bibitem[{Tang and Ng(2006)}]{tang2006googling}
\bibinfo{author}{Tang\xfnm[ H.]}, \bibinfo{author}{Ng\xfnm[ J.H.K.]}.
\newblock \bibinfo{title}{Googling for a diagnosis—use of google as a
  diagnostic aid: internet based study}.
\newblock \bibinfo{journal}{Bmj}
  \bibinfo{year}{2006};\bibinfo{volume}{333}(\bibinfo{number}{7579}):\bibinfo{pages}{1143--1145}.
\bibitem[{Falagas et~al.(2009)Falagas, Ntziora, Makris, Malietzis and
  Rafailidis}]{falagas2009pubmed}
\bibinfo{author}{Falagas\xfnm[ M.E.]}, \bibinfo{author}{Ntziora\xfnm[ F.]},
  \bibinfo{author}{Makris\xfnm[ G.C.]}, \bibinfo{author}{Malietzis\xfnm[
  G.A.]}, \bibinfo{author}{Rafailidis\xfnm[ P.I.]}.
\newblock \bibinfo{title}{Do pubmed and google searches help medical students
  and young doctors reach the correct diagnosis? a pilot study}.
\newblock \bibinfo{journal}{European journal of internal medicine}
  \bibinfo{year}{2009};\bibinfo{volume}{20}(\bibinfo{number}{8}):\bibinfo{pages}{788--790}.
\bibitem[{Walji et~al.(2005)Walji, Sagaram, Meric-Bernstam, Johnson and
  Bernstam}]{walji2005searching}
\bibinfo{author}{Walji\xfnm[ M.]}, \bibinfo{author}{Sagaram\xfnm[ S.]},
  \bibinfo{author}{Meric-Bernstam\xfnm[ F.]}, \bibinfo{author}{Johnson\xfnm[
  C.W.]}, \bibinfo{author}{Bernstam\xfnm[ E.V.]}.
\newblock \bibinfo{title}{Searching for cancer-related information online:
  Unintended retrieval of complementary and alternative medicine information}.
\newblock \bibinfo{journal}{International journal of medical informatics}
  \bibinfo{year}{2005};\bibinfo{volume}{74}(\bibinfo{number}{7}):\bibinfo{pages}{685--693}.
\bibitem[{Croft et~al.(2010)Croft, Metzler and Strohman}]{croft2010search}
\bibinfo{author}{Croft\xfnm[ W.B.]}, \bibinfo{author}{Metzler\xfnm[ D.]},
  \bibinfo{author}{Strohman\xfnm[ T.]}.
\newblock \bibinfo{title}{Search engines: Information retrieval in practice}.
\newblock \bibinfo{publisher}{Addison-Wesley}; \bibinfo{year}{2010}.
\bibitem[{Buckley and Voorhees(2004)}]{buckley2004retrieval}
\bibinfo{author}{Buckley\xfnm[ C.]}, \bibinfo{author}{Voorhees\xfnm[ E.M.]}.
\newblock \bibinfo{title}{Retrieval evaluation with incomplete information}.
\newblock In: \bibinfo{booktitle}{Proceedings of the 27th annual international
  ACM SIGIR conference on Research and development in information retrieval}.
  \bibinfo{organization}{ACM}; \bibinfo{year}{2004}, p.
  \bibinfo{pages}{25--32}.
\bibitem[{Voorhees and Tice(1999)}]{voorhees1999trec}
\bibinfo{author}{Voorhees\xfnm[ E.]}, \bibinfo{author}{Tice\xfnm[ D.M.]}.
\newblock \bibinfo{title}{The trec-8 question answering track evaluation}.
\newblock In: \bibinfo{booktitle}{Proceedings of The Eighth Text REtrieval
  Conference (TREC-8), http://trec. nist. gov/pubs/trec8/t8\_proceedings.
  html}. \bibinfo{year}{1999},.
\bibitem[{Haran and Schattner(2011)}]{haran2011clinical}
\bibinfo{author}{Haran\xfnm[ M.]}, \bibinfo{author}{Schattner\xfnm[ A.]}.
\newblock \bibinfo{title}{On the clinical encounter with'zebras'--the science
  and art of rare diseases}.
\newblock \bibinfo{journal}{European journal of internal medicine}
  \bibinfo{year}{2011};\bibinfo{volume}{22}(\bibinfo{number}{3}):\bibinfo{pages}{235--237}.
\bibitem[{Strohman et~al.(2005)Strohman, Metzler, Turtle and
  Croft}]{strohman2005indri}
\bibinfo{author}{Strohman\xfnm[ T.]}, \bibinfo{author}{Metzler\xfnm[ D.]},
  \bibinfo{author}{Turtle\xfnm[ H.]}, \bibinfo{author}{Croft\xfnm[ W.B.]}.
\newblock \bibinfo{title}{Indri: A language model-based search engine for
  complex queries}.
\newblock In: \bibinfo{booktitle}{Proceedings of the International Conference
  on Intelligent Analysis}; vol.~\bibinfo{volume}{2}.
  \bibinfo{organization}{Citeseer}; \bibinfo{year}{2005}, p.
  \bibinfo{pages}{2--6}.
\bibitem[{Croft and Lafferty(2003)}]{croft2003language}
\bibinfo{author}{Croft\xfnm[ W.B.]}, \bibinfo{author}{Lafferty\xfnm[ J.]}.
\newblock \bibinfo{title}{Language modeling for information retrieval};
  vol.~\bibinfo{volume}{13}.
\newblock \bibinfo{publisher}{Springer}; \bibinfo{year}{2003}.
\bibitem[{Zhai and Lafferty(2004)}]{zhai2004study}
\bibinfo{author}{Zhai\xfnm[ C.]}, \bibinfo{author}{Lafferty\xfnm[ J.]}.
\newblock \bibinfo{title}{A study of smoothing methods for language models
  applied to information retrieval}.
\newblock \bibinfo{journal}{ACM Transactions on Information Systems (TOIS)}
  \bibinfo{year}{2004};\bibinfo{volume}{22}(\bibinfo{number}{2}):\bibinfo{pages}{179--214}.
\bibitem[{Voorhees and Buckland(2011)}]{voorhees2011trec}
\bibinfo{editor}{Voorhees\xfnm[ E.M.]}, \bibinfo{editor}{Buckland\xfnm[ L.P.]},
  editors.
\newblock \bibinfo{title}{Proceedings of The Twentieth Text REtrieval
  Conference, TREC 2011, Gaithersburg, Maryland, November 15-18, 2011}.
  \bibinfo{publisher}{National Institute of Standards and Technology (NIST)};
  \bibinfo{year}{2011}.
\bibitem[{De~Somer et~al.(2010)De~Somer, Wouters, Morren, De~Vos, Van Den~Oord,
  Devriendt et~al.}]{de2010granulomatous}
\bibinfo{author}{De~Somer\xfnm[ L.]}, \bibinfo{author}{Wouters\xfnm[ C.]},
  \bibinfo{author}{Morren\xfnm[ M.A.]}, \bibinfo{author}{De~Vos\xfnm[ R.]},
  \bibinfo{author}{Van Den~Oord\xfnm[ J.]}, \bibinfo{author}{Devriendt\xfnm[
  K.]}, et~al.
\newblock \bibinfo{title}{Granulomatous skin lesions complicating varicella
  infection in a patient with rothmund-thomson syndrome and immune deficiency:
  case report}.
\newblock \bibinfo{journal}{Orphanet Journal of Rare Diseases}
  \bibinfo{year}{2010};\bibinfo{volume}{5}(\bibinfo{number}{1}):\bibinfo{pages}{37}.
\bibitem[{B{\"o}hm et~al.(2010)B{\"o}hm, Yi{\c{s}}, Orta{\c{c}},
  {\c{C}}akmak{\c{c}}{\i}, Kurul, Dirik et~al.}]{bohm2010case}
\bibinfo{author}{B{\"o}hm\xfnm[ J.]}, \bibinfo{author}{Yi{\c{s}}\xfnm[ U.]},
  \bibinfo{author}{Orta{\c{c}}\xfnm[ R.]},
  \bibinfo{author}{{\c{C}}akmak{\c{c}}{\i}\xfnm[ H.]},
  \bibinfo{author}{Kurul\xfnm[ S.H.]}, \bibinfo{author}{Dirik\xfnm[ E.]},
  et~al.
\newblock \bibinfo{title}{Case report of intrafamilial variability in autosomal
  recessive centronuclear myopathy associated to a novel bin1 stop mutation}.
\newblock \bibinfo{journal}{Orphanet J Rare Dis}
  \bibinfo{year}{2010};\bibinfo{volume}{5}:\bibinfo{pages}{35}.
\bibitem[{Schneider et~al.(2010)Schneider, Lingesleben, Vogel, Garuti and
  Calandra}]{schneider2010novel}
\bibinfo{author}{Schneider\xfnm[ H.]}, \bibinfo{author}{Lingesleben\xfnm[ A.]},
  \bibinfo{author}{Vogel\xfnm[ H.P.]}, \bibinfo{author}{Garuti\xfnm[ R.]},
  \bibinfo{author}{Calandra\xfnm[ S.]}.
\newblock \bibinfo{title}{A novel mutation in the sterol 27-hydroxylase gene of
  a woman with autosomal recessive cerebrotendinous xanthomatosis}.
\newblock \bibinfo{journal}{Orphanet journal of rare diseases}
  \bibinfo{year}{2010};\bibinfo{volume}{5}(\bibinfo{number}{1}):\bibinfo{pages}{27}.
\bibitem[{Orsoni et~al.(2010)Orsoni, Lagan{\`a}, Rubino, Zavota, Bacciu and
  Mora}]{orsoni2010case}
\bibinfo{author}{Orsoni\xfnm[ J.G.]}, \bibinfo{author}{Lagan{\`a}\xfnm[ B.]},
  \bibinfo{author}{Rubino\xfnm[ P.]}, \bibinfo{author}{Zavota\xfnm[ L.]},
  \bibinfo{author}{Bacciu\xfnm[ S.]}, \bibinfo{author}{Mora\xfnm[ P.]}.
\newblock \bibinfo{title}{Case report rituximab ameliorated severe hearing loss
  in cogan's syndrome: a case report}.
\newblock \bibinfo{journal}{Orphanet Journal of Rare Diseases}
  \bibinfo{year}{2010};.
\bibitem[{Al-Owain et~al.(2010)Al-Owain, Mohamed, Kaya, Zagal, Matthijs and
  Jaeken}]{al2010case}
\bibinfo{author}{Al-Owain\xfnm[ M.]}, \bibinfo{author}{Mohamed\xfnm[ S.]},
  \bibinfo{author}{Kaya\xfnm[ N.]}, \bibinfo{author}{Zagal\xfnm[ A.]},
  \bibinfo{author}{Matthijs\xfnm[ G.]}, \bibinfo{author}{Jaeken\xfnm[ J.]}.
\newblock \bibinfo{title}{Case report a novel mutation and first report of
  dilated cardiomyopathy in alg6-cdg (cdg-ic): a case report}.
\newblock \bibinfo{journal}{Orphanet Journal of Rare Diseases}
  \bibinfo{year}{2010};.
\bibitem[{Aci{\'e}n et~al.(2010)Aci{\'e}n, Gal{\'a}n, Manch{\'o}n, Ruiz,
  Aci{\'e}n and Alcaraz}]{acien2010hereditary}
\bibinfo{author}{Aci{\'e}n\xfnm[ P.]}, \bibinfo{author}{Gal{\'a}n\xfnm[ F.]},
  \bibinfo{author}{Manch{\'o}n\xfnm[ I.]}, \bibinfo{author}{Ruiz\xfnm[ E.]},
  \bibinfo{author}{Aci{\'e}n\xfnm[ M.]}, \bibinfo{author}{Alcaraz\xfnm[ L.A.]}.
\newblock \bibinfo{title}{Hereditary renal adysplasia, pulmonary hypoplasia and
  mayer-rokitansky-k{\"u}ster-hauser (mrkh) syndrome: a case report}.
\newblock \bibinfo{journal}{Orphanet Journal of Rare Diseases}
  \bibinfo{year}{2010};\bibinfo{volume}{5}:\bibinfo{pages}{6}.
\bibitem[{Poala et~al.(2010)Poala, Bisogno and Colombatti}]{poala2010case}
\bibinfo{author}{Poala\xfnm[ S.B.]}, \bibinfo{author}{Bisogno\xfnm[ G.]},
  \bibinfo{author}{Colombatti\xfnm[ R.]}.
\newblock \bibinfo{title}{Case report thrombocytopenia and splenomegaly: an
  unusual presentation of congenital hepatic fibrosis}.
\newblock \bibinfo{journal}{Orphanet Journal of Rare Diseases}
  \bibinfo{year}{2010};.
\bibitem[{Mejia-Gaviria et~al.(2010)Mejia-Gaviria, Gil-Pe{\~n}a, Coto,
  P{\'e}rez-Men{\'e}ndez, Santos et~al.}]{mejia2010genetic}
\bibinfo{author}{Mejia-Gaviria\xfnm[ N.]}, \bibinfo{author}{Gil-Pe{\~n}a\xfnm[
  H.]}, \bibinfo{author}{Coto\xfnm[ E.]},
  \bibinfo{author}{P{\'e}rez-Men{\'e}ndez\xfnm[ T.M.]},
  \bibinfo{author}{Santos\xfnm[ F.]}, et~al.
\newblock \bibinfo{title}{Genetic and clinical peculiarities in a new family
  with hereditary hypophosphatemic rickets with hypercalciuria: a case report}.
\newblock \bibinfo{journal}{Orphanet J Rare Dis}
  \bibinfo{year}{2010};\bibinfo{volume}{5}(\bibinfo{number}{1}):\bibinfo{pages}{1}.
\bibitem[{Cassiman et~al.(2009)Cassiman, Zeevaert, Holme, Kvittingen, Jaeken
  et~al.}]{cassiman2009novel}
\bibinfo{author}{Cassiman\xfnm[ D.]}, \bibinfo{author}{Zeevaert\xfnm[ R.]},
  \bibinfo{author}{Holme\xfnm[ E.]}, \bibinfo{author}{Kvittingen\xfnm[ E.A.]},
  \bibinfo{author}{Jaeken\xfnm[ J.]}, et~al.
\newblock \bibinfo{title}{A novel mutation causing mild, atypical
  fumarylacetoacetase deficiency (tyrosinemia type i): a case report}.
\newblock \bibinfo{journal}{Orphanet journal of rare diseases}
  \bibinfo{year}{2009};\bibinfo{volume}{4}(\bibinfo{number}{1}):\bibinfo{pages}{28}.
\bibitem[{Drera et~al.(2009)Drera, Ritelli, Zoppi, Wischmeijer, Gnoli, Fattori
  et~al.}]{drera2009loeys}
\bibinfo{author}{Drera\xfnm[ B.]}, \bibinfo{author}{Ritelli\xfnm[ M.]},
  \bibinfo{author}{Zoppi\xfnm[ N.]}, \bibinfo{author}{Wischmeijer\xfnm[ A.]},
  \bibinfo{author}{Gnoli\xfnm[ M.]}, \bibinfo{author}{Fattori\xfnm[ R.]},
  et~al.
\newblock \bibinfo{title}{Loeys-dietz syndrome type i and type ii: clinical
  findings and novel mutations in two italian patients}.
\newblock \bibinfo{journal}{Orphanet J Rare Dis}
  \bibinfo{year}{2009};\bibinfo{volume}{4}:\bibinfo{pages}{24}.
\bibitem[{Ritelli et~al.(2009)Ritelli, Drera, Vicchio, Puppini, Biban, Pilati
  et~al.}]{ritelli2009arterial}
\bibinfo{author}{Ritelli\xfnm[ M.]}, \bibinfo{author}{Drera\xfnm[ B.]},
  \bibinfo{author}{Vicchio\xfnm[ M.]}, \bibinfo{author}{Puppini\xfnm[ G.]},
  \bibinfo{author}{Biban\xfnm[ P.]}, \bibinfo{author}{Pilati\xfnm[ M.]}, et~al.
\newblock \bibinfo{title}{Arterial tortuosity syndrome in two italian
  paediatric patients}.
\newblock \bibinfo{journal}{Orphanet J Rare Dis}
  \bibinfo{year}{2009};\bibinfo{volume}{4}:\bibinfo{pages}{20}.
\bibitem[{Spoerl et~al.(2009)Spoerl, B{\"a}r, Cooper, Vogt, Tyndall and
  Walker}]{spoerl2009multisegmental}
\bibinfo{author}{Spoerl\xfnm[ D.]}, \bibinfo{author}{B{\"a}r\xfnm[ D.]},
  \bibinfo{author}{Cooper\xfnm[ J.]}, \bibinfo{author}{Vogt\xfnm[ T.]},
  \bibinfo{author}{Tyndall\xfnm[ A.]}, \bibinfo{author}{Walker\xfnm[ U.A.]}.
\newblock \bibinfo{title}{Multisegmental spondylitis due to tropheryma
  whipplei: Case report}.
\newblock \bibinfo{journal}{Orphanet Journal of Rare Diseases}
  \bibinfo{year}{2009};\bibinfo{volume}{4}(\bibinfo{number}{1}):\bibinfo{pages}{13}.
\bibitem[{Marchiori et~al.(2009)Marchiori, Ferreira, Bittencourt,
  de~Ara{\'u}jo~Neto, Zanetti, Mano et~al.}]{marchiori2009pulmonary}
\bibinfo{author}{Marchiori\xfnm[ E.]}, \bibinfo{author}{Ferreira\xfnm[ J.]},
  \bibinfo{author}{Bittencourt\xfnm[ C.N.]},
  \bibinfo{author}{de~Ara{\'u}jo~Neto\xfnm[ C.A.]},
  \bibinfo{author}{Zanetti\xfnm[ G.]}, \bibinfo{author}{Mano\xfnm[ C.M.]},
  et~al.
\newblock \bibinfo{title}{Pulmonary hemorrhage syndrome associated with dengue
  fever, highresolution computed tomography findings: a case report}.
\newblock \bibinfo{journal}{Orphanet J Rare Dis}
  \bibinfo{year}{2009};\bibinfo{volume}{4}(\bibinfo{number}{8}).
\bibitem[{Zamel et~al.(2008)Zamel, Khan, Pollex, Hegele
  et~al.}]{zamel2008abetalipoproteinemia}
\bibinfo{author}{Zamel\xfnm[ R.]}, \bibinfo{author}{Khan\xfnm[ R.]},
  \bibinfo{author}{Pollex\xfnm[ R.L.]}, \bibinfo{author}{Hegele\xfnm[ R.A.]},
  et~al.
\newblock \bibinfo{title}{Abetalipoproteinemia: two case reports and literature
  review}.
\newblock \bibinfo{journal}{Orphanet J Rare Dis}
  \bibinfo{year}{2008};\bibinfo{volume}{3}:\bibinfo{pages}{19}.
\bibitem[{Virginie et~al.(????)Virginie, Isabelle, Alain, C{\'e}cile, Christine
  and Nathalie}]{virginie3carglumic}
\bibinfo{author}{Virginie\xfnm[ L.]}, \bibinfo{author}{Isabelle\xfnm[ F.]},
  \bibinfo{author}{Alain\xfnm[ F.]}, \bibinfo{author}{C{\'e}cile\xfnm[ A.]},
  \bibinfo{author}{Christine\xfnm[ V.S.]}, \bibinfo{author}{Nathalie\xfnm[
  G.]}.
\newblock \bibinfo{title}{Carglumic acid: an additional therapy in the
  treatment of organic acidurias with hyperammonemia?}
\newblock \bibinfo{journal}{Orphanet Journal of Rare Diseases}
  ????;\bibinfo{volume}{3}.
\bibitem[{Joy et~al.(2007)Joy, Cao, Black, Malik, Charlton-Menys, Hegele
  et~al.}]{joy2007alstrom}
\bibinfo{author}{Joy\xfnm[ T.]}, \bibinfo{author}{Cao\xfnm[ H.]},
  \bibinfo{author}{Black\xfnm[ G.]}, \bibinfo{author}{Malik\xfnm[ R.]},
  \bibinfo{author}{Charlton-Menys\xfnm[ V.]}, \bibinfo{author}{Hegele\xfnm[
  R.A.]}, et~al.
\newblock \bibinfo{title}{Alstrom syndrome (omim 203800): a case report and
  literature review}.
\newblock \bibinfo{journal}{Orphanet Journal of Rare Diseases}
  \bibinfo{year}{2007};\bibinfo{volume}{2}(\bibinfo{number}{1}):\bibinfo{pages}{49}.
\bibitem[{Ceruti et~al.(2007)Ceruti, Rodi, Stella, Adami, Bolongaro, Baritussio
  et~al.}]{ceruti2007successful}
\bibinfo{author}{Ceruti\xfnm[ M.]}, \bibinfo{author}{Rodi\xfnm[ G.]},
  \bibinfo{author}{Stella\xfnm[ G.M.]}, \bibinfo{author}{Adami\xfnm[ A.]},
  \bibinfo{author}{Bolongaro\xfnm[ A.]}, \bibinfo{author}{Baritussio\xfnm[
  A.]}, et~al.
\newblock \bibinfo{title}{Successful whole lung lavage in pulmonary alveolar
  proteinosis secondary to lysinuric protein intolerance: a case report}.
\newblock \bibinfo{journal}{Orphanet J Rare Dis}
  \bibinfo{year}{2007};\bibinfo{volume}{2}:\bibinfo{pages}{14}.
\bibitem[{Gruhn et~al.(2007)Gruhn, Seidel, Zintl, Varon, T{\"o}nnies, Neitzel
  et~al.}]{gruhn2007successful}
\bibinfo{author}{Gruhn\xfnm[ B.]}, \bibinfo{author}{Seidel\xfnm[ J.]},
  \bibinfo{author}{Zintl\xfnm[ F.]}, \bibinfo{author}{Varon\xfnm[ R.]},
  \bibinfo{author}{T{\"o}nnies\xfnm[ H.]}, \bibinfo{author}{Neitzel\xfnm[ H.]},
  et~al.
\newblock \bibinfo{title}{Successful bone marrow transplantation in a patient
  with dna ligase iv deficiency and bone marrow failure}.
\newblock \bibinfo{journal}{Orphanet J Rare Dis}
  \bibinfo{year}{2007};\bibinfo{volume}{2}(\bibinfo{number}{5}).
\bibitem[{Al~Kaissi et~al.(2007)Al~Kaissi, Grill, Safi, Ben~Ghachem,
  Ben~Chehida and Klaushofer}]{al2007craniocervical}
\bibinfo{author}{Al~Kaissi\xfnm[ A.]}, \bibinfo{author}{Grill\xfnm[ F.]},
  \bibinfo{author}{Safi\xfnm[ H.]}, \bibinfo{author}{Ben~Ghachem\xfnm[ M.]},
  \bibinfo{author}{Ben~Chehida\xfnm[ F.]}, \bibinfo{author}{Klaushofer\xfnm[
  K.]}.
\newblock \bibinfo{title}{Craniocervical junction malformation in a child with
  oromandibular-limb hypogenesis-m{\"o}bius syndrome}.
\newblock \bibinfo{journal}{Orphanet J Rare Dis}
  \bibinfo{year}{2007};\bibinfo{volume}{8}(\bibinfo{number}{2}):\bibinfo{pages}{2}.
\bibitem[{Sills et~al.(2007)Sills, Burns, Parker, Carroll, Kephart, Dyer
  et~al.}]{sills2007further}
\bibinfo{author}{Sills\xfnm[ E.S.]}, \bibinfo{author}{Burns\xfnm[ M.]},
  \bibinfo{author}{Parker\xfnm[ L.D.]}, \bibinfo{author}{Carroll\xfnm[ L.P.]},
  \bibinfo{author}{Kephart\xfnm[ L.L.]}, \bibinfo{author}{Dyer\xfnm[ C.]},
  et~al.
\newblock \bibinfo{title}{Further phenotypic delineation of subtelomeric
  (terminal) 4q deletion with emphasis on intracranial and reproductive
  anatomy}.
\newblock \bibinfo{journal}{Orphanet Journal of Rare Diseases}
  \bibinfo{year}{2007};\bibinfo{volume}{2}(\bibinfo{number}{1}):\bibinfo{pages}{9}.

\end{thebibliography}

\end{document}